\documentclass[12pt]{article}

\usepackage{fullpage}
\usepackage{amssymb}
\usepackage{bbm}
\usepackage{color}
\usepackage{ulem}
\usepackage[numbers,sort&compress]{natbib}

\newcommand{\Comment}[1]{{}}
\definecolor{purple}{rgb}{0.4,0.1,0.55}
\usepackage[linktocpage=true]{hyperref}
\hypersetup{
colorlinks=true,
citecolor=purple,
linkcolor=purple,
urlcolor=purple,
pdfauthor={Goon, Hinterbichler, Joyce and Trodden},
pdftitle={Galileons as Wess--Zumino Terms},
pdfsubject={}
}

\DeclareFontFamily{OT1}{rsfs10}{}
\DeclareFontShape{OT1}{rsfs10}{m}{n}{ <-> rsfs10 }{}
\DeclareMathAlphabet{\mathscript}{OT1}{rsfs10}{m}{n}

\def\gsim{ \lower .75ex \hbox{$\sim$} \llap{\raise .27ex \hbox{$>$}} }
\def\lsim{ \lower .75ex \hbox{$\sim$} \llap{\raise .27ex \hbox{$<$}} }
\def\be{\begin{equation}}
\def\ee{\end{equation}}
\def\bea{\begin{eqnarray}}
\def\eea{\end{eqnarray}}

\newcommand{\ns}{\normalsize}

\newcommand{\half}{\frac{1}{2}}

\def\mn{_{\mu \nu}}

\newcommand{\rd}{{\rm d}}

\def\({\left(}
\def\){\right)}

\usepackage{latexsym,amsmath,amssymb,epsfig}

\usepackage{graphicx}
\usepackage{graphicx,subfigure}
\usepackage{epstopdf}
\usepackage[body={17.5cm, 21cm},right=2cm]{geometry}
\usepackage{amssymb}
\usepackage{amsmath}
\usepackage{psfrag}
\usepackage{epsfig}
\usepackage{cancel}
 \allowdisplaybreaks[4]

\usepackage[all]{xy}

\begin{document}

\begin{titlepage}

\title{
  \hfill{\ns }  \\[1em]
   {\LARGE\sc Galileons as Wess--Zumino Terms}
\\[1em] }
\author{\ns\large
  Garrett Goon, Kurt Hinterbichler, Austin Joyce and Mark Trodden
     \\[0.5em]
{\ns  Center for Particle Cosmology, Department of Physics and Astronomy,} \\[-0.3cm]
{\ns  University of Pennsylvania, Philadelphia, PA 19104}\\[0.3cm]
}
\date{}
\maketitle

\begin{abstract}
We show that the galileons can be thought of as Wess--Zumino terms for the spontaneous breaking of space-time symmetries. Wess--Zumino terms are terms which are not captured by the coset construction for phenomenological Lagrangians with broken symmetries.  Rather they are, in $d$ space-time dimensions, $d$-form potentials for $(d+1)$-forms which are non-trivial co-cycles in Lie algebra cohomology of the full symmetry group relative to the unbroken symmetry group. We introduce the galileon algebras and construct the non-trivial $(d+1)$-form co-cycles, showing that the presence of galileons and multi-galileons in all dimensions is counted by the dimensions of particular Lie algebra cohomology groups. We also discuss the DBI and conformal galileons from this point of view, showing that they are not Wess--Zumino terms, with one exception in each case.
\end{abstract}

\end{titlepage}
\setcounter{page}{2}

\tableofcontents

\section{Introduction}
\numberwithin{equation}{section}
The study of higher-dimensional origins for consistent infrared modifications of gravity has led to the discovery of novel four-dimensional scalar field theories with intriguing properties, which point to interesting implications for both particle physics and cosmology. The simplest, and original example is provided by the Dvali--Gabadadze--Porrati (DGP) model~\cite{Dvali:2000hr}, which describes a 3-brane floating in a five-dimensional bulk via an action containing both a bulk and brane Einstein--Hilbert term. It is possible to write a four-dimensional effective action for this model and to take a decoupling limit, in which Einstein gravity is modified by the presence of an additional scalar, $\pi$, which possesses an interaction of the form $\sim \square\pi(\partial\pi)^2$ \cite{Luty:2003vm}.

Though this interaction is higher-derivative, it nevertheless has second order equations of motion. This guarantees that the theory does not propagate a ghost, which is the usual pathology associated with many higher-derivative scalars. From the higher-dimensional viewpoint, the $\pi$ field is the brane-bending mode---the Goldstone field associated with spontaneously broken five-dimensional Poincar\'e invariance. In a certain limit, this non-linearly realized symmetry manifests itself as a ``galilean" shift symmetry of the scalar 
\be
\pi(x)\longrightarrow\pi(x)+c+b_\mu x^\mu~. \label{shiftsym}
\ee
Although terms of this type have their origins in brane-world modified gravity models, they have since been abstracted, and studied in their own right~\cite{Nicolis:2008in}, with the relevant scalar field named the {\it galileon} (for a review of recent developments, see~\cite{Trodden:2011xh}). 

In four dimensions, it is possible to construct five terms which have both second-order equations of motion and are invariant under this galilean shift symmetry.  In $d$ dimensions, there are $d+1$ such terms.  For $1\leq n\leq d$, the $n$-th order galileon Lagrangian is
\be
\label{galileon0} 
{\cal L}_{n}\sim \eta^{\mu_1\nu_1\mu_2\nu_2\cdots\mu_{n-1}\nu_{n-1}}\left( \pi\partial_{\mu_1}\partial_{\nu_1}\pi\partial_{\mu_2}\partial_{\nu_2}\pi\cdots\partial_{\mu_{n-1}}\partial_{\nu_{n-1}}\pi\right),
\ee 
where $\eta^{\mu_1\nu_1\mu_2\nu_2\cdots\mu_n\nu_n}\equiv{1\over n!}\sum_p\left(-1\right)^{p}\eta^{\mu_1p(\nu_1)}\eta^{\mu_2p(\nu_2)}\cdots\eta^{\mu_np(\nu_n)}$, 
the sum over all permutations of the $\nu$ indices, with $(-1)^p$ the sign of the permutation.  The first is a tadpole,
 ${\cal L}_1\sim \pi,$ the second is the kinetic term ${\cal L}_2\sim (\partial\pi)^2$, and the third ${\cal L}_3\sim \square\pi(\partial\pi)^2$ is the cubic DGP-like term. 

It is also possible to construct ${\rm SO}(N)$ symmetric multi-galileon theories, where the fields $\pi^I$ each have the shift symmetry~(\ref{shiftsym}) and also rotate in the fundamental representation of an internal ${\rm SO}(N)$~\cite{Padilla:2010de,Hinterbichler:2010xn}.  In this case, in $d$ dimensions there are $d/2$ possible galileon terms if $d$ is even, and $(d+1)/2$ if $d$ is odd.  Only galileons for $n$ even exist, containing an even number of $\pi^I$'s (thus, there is no tadpole). These are obtained by simply contracting indices with $\delta_{IJ}$,
\begin{align}
\label{multigalileon0} 
\nonumber
{\cal L}_{n}\sim~&
\delta_{I_1J_1}\delta_{I_2J_2}\cdots \delta_{I_{n/2}J_{n/2}}  \eta^{\mu_1\nu_1\mu_2\nu_2\cdots\mu_{n-1}\nu_{n-1}}\\
&\times \left(\pi^{I_1}\partial_{\mu_1}\partial_{\nu_1}\pi^{J_1}\partial_{\mu_2}\partial_{\nu_2}\pi^{I_2}\partial_{\mu_3}\partial_{\nu_3}\pi^{J_2}
\cdots\partial_{\mu_{n-2}}\partial_{\nu_{n-2}}\pi^{I_{n/2}}\partial_{\mu_{n-1}}\partial_{\nu_{n-1}}\pi^{J_{n/2}}\right). 
\end{align}

There are two further important properties for what we will have to say.  First, the galileon terms are not strictly invariant under the symmetry (\ref{shiftsym}), but rather shift by a total derivative, leaving the action invariant.  Second, the $n$-th galileon has $2n-2$ derivatives, so they have fewer than two derivatives per field, whereas every other possible term invariant under  (\ref{shiftsym}) has at least two derivatives per field.

Much of the interest in galileons is due to their attractive field-theoretic properties. The fact that they have fewer derivatives than other terms invariant under the shift symmetry makes it possible to find regimes in which the galileons can be consistently treated as the only important interactions~\cite{Nicolis:2004qq}.  Furthermore, around sources,  galileon theories exhibit the Vainshtein screening mechanism~\cite{Vainshtein:1972sx, Deffayet:2001uk} at short distances, allowing them to evade fifth force constraints, such as those provided by measurements within the solar system. Finally, the galileon terms are not renormalized to any loop order in perturbation theory~\cite{Luty:2003vm, Hinterbichler:2010xn}, allowing them to be treated classically.

Theories of this type have been used for many phenomenological applications in both the early and late universe, including inflation~\cite{Kobayashi:2010cm, Burrage:2010cu, Creminelli:2010qf}, alternatives to inflation~\cite{Creminelli:2010ba,Hinterbichler:2011qk,Hinterbichler:2012mv}, and late-time cosmic acceleration~\cite{Chow:2009fm, Silva:2009km, DeFelice:2010as, Deffayet:2010qz}. They have been covariantized and coupled to gravity~\cite{Deffayet:2009wt, Deffayet:2009mn} as well as extended to $p$-forms~\cite{Deffayet:2010zh}, supersymmetrized \cite{Khoury:2011da} and coupled to gauge fields~\cite{Zhou:2011ix, Goon:2012mu}. Galileons also appear in the scalar sector of ghost-free massive gravity~\cite{deRham:2010ik,deRham:2010kj} (for a review, see~\cite{Hinterbichler:2011tt}).

The construction of galileon theories can be illuminating itself.  One instructive method of deriving the galileon terms is via the probe brane construction of~\cite{deRham:2010eu}, in which a 3-brane probes a five-dimensional bulk. From this geometric perspective, galileon terms appear as the small-field limit of Lovelock invariants of the induced brane metric and from Gibbons--Hawking--York boundary terms associated with bulk Lovelock invariants. The appearance of Lovelock invariants sheds some light on the fact that galileon terms have second order equations of motion---Lovelock terms are the only terms that may be added to Einstein gravity while maintaining second order equations of motion for the metric~\cite{Lovelock:1971yv}. The probe brane construction has been extended in various directions, most notably to higher co-dimension~\cite{Hinterbichler:2010xn, Goon:2012mu}---leading to the multi-galileons with an internal global SO$(N)$ symmetry among the fields (which can furthermore be gauged \cite{Zhou:2011ix,Goon:2012mu})---and to curved backgrounds~\cite{Goon:2011qf,Goon:2011uw, Burrage:2011bt, Goon:2011xf}, where the fields are invariant under complicated non-linear symmetries inherited from bulk Killing vectors.

In this paper we present a different method of deriving the galileon terms---an algebraic method, treating them as Goldstone modes of spontaneously broken space-time symmetries.  We employ the techniques of non-linear realizations developed by Callan, Coleman, Wess and Zumino~\cite{Coleman:1969sm, Callan:1969sn} and Volkov \cite{volkov}. We show that, like the familiar Wess--Zumino--Witten term of the chiral Lagrangian \cite{Wess:1971yu,Witten:1983tw}, the galileon terms in $d$ dimensions are not captured by the na\"ive $d$-dimensional coset construction. Instead, the galileons arise from invariant $(d+1)$-forms created via the coset construction which are then pulled back to our $d$ dimensional space-time in order to create galileon invariant actions.  The relevant $(d+1)$-forms, and hence the galileons, are associated with non-trivial co-cycles in an appropriate Lie algebra cohomology~\cite{Chevalley:1948zz,deAzcarraga:1998uy,de Azcarraga:1997gn}, which is a cohomology theory on forms which are left-invariant under vector fields that generate the symmetry algebra.\footnote{A similar viewpoint was conveyed in~\cite{Brugues:2004an}, where the low-energy effective actions for non-relativistic strings and branes were obtained as Wess--Zumino terms.} This is related to the internal symmetry case, where it was shown in \cite{D'Hoker:1994ti} that Wess--Zumino terms are counted by de Rham cohomology. Indeed, for compact groups, de Rham and Lie algebra cohomology are isomorphic \cite{deAzcarraga:1995jw}.

After reviewing the general coset construction, we describe the algebra non-linearly realized by the galileons---the ``galileon algebra."  We show that, inspired by brane-world models, this is a contraction of a higher-dimensional Poincar\'e algebra only along particular auxiliary directions, that is, it can be thought of as the Poincar\'e algebra of a brane embedded in higher dimensions, where the speed of light in the directions transverse to the brane is sent to infinity, while the speed of light along the brane is kept constant. The most familiar example of a galileon theory is the non-relativistic free point particle, which can be thought of as a $(0+1)$-dimensional field theory invariant under the galilean group.  We review the construction of the kinetic term for the free particle as a Wess--Zumino term before applying our arguments to the most physically relevant situation of galileons in four dimensions. As the galileons are Wess--Zumino terms, we argue that the number of such terms for both single and multi-galileon situations is bounded by the dimension of the appropriate Lie algebra cohomology groups.

Additionally, we consider the conformal galileons.  In this case, only one of the conformal galileons, the cubic term, appears as a Wess--Zumino term for spontaneously broken conformal symmetry.  We construct this Wess--Zumino term explicitly and comment on its relation to the curvature invariant technique employed in \cite{Nicolis:2008in} to construct the conformal galileons.

Finally, we demonstrate that, although the original galileons are Wess--Zumino terms for spontaneously broken space-time symmetries, this is not the case for the relativistic DBI galileons \cite{deRham:2010eu,Goon:2010xh}, which---aside from the tadpole term---are obtainable from the coset construction and hence are not Wess--Zumino terms. We show how to construct the DBI galileons using the techniques of non-linear realizations.

{\textbf{Conventions:}} We use the mostly plus metric convention.  The number of spacetime dimensions is denoted by $d$.  The flat space epsilon tensor is defined so that $\epsilon_{01\cdots d}=+1$.  Indices are anti-symmetrized with weight one.
\section{Nonlinear realizations and the coset construction}
The galileon actions are invariant under the non-linear symmetries (\ref{shiftsym}), and may therefore be interpreted as Goldstone bosons arising from spontaneous symmetry breaking.  Broken symmetries and effective field theory have historically been extremely profitable viewpoints from which to study the low-energy dynamics of physical systems. Motivated by the successes of phenomenological Lagrangians in describing low energy pion scattering \cite{Weinberg:1968de}, Callan, Coleman, Wess and Zumino \cite{Coleman:1969sm, Callan:1969sn}, as well as Volkov \cite{volkov}, developed a powerful formalism for constructing the most general effective action for a given symmetry breaking pattern. This is the now well-known technique of non-linear realizations, or coset construction, which we review briefly here. More comprehensive reviews are given in \cite{xthschool, Zumino:1970tu}.

\subsection{Spontaneously broken internal symmetries\label{spontbrokeninternalsymmetries}}

We begin by reviewing the problem of constructing a Lagrangian for Goldstone fields corresponding to the breaking of an internal ({\it i.e.}, commuting with the Poincar\'e group) symmetry group $G$ down to a subgroup $H$; that is, we seek the most general Lagrangian which is invariant under $G$ transformations, where the $H$ transformations act linearly on the fields and those not in $H$ act non-linearly.  As is well known \cite{Coleman:1969sm, Callan:1969sn}, there will be ${\rm dim}({G/H})$ Goldstone bosons, which parametrize the space of (left) cosets ${ G/H}$. 

However, to start with, we use fields $V(x)$ that take values in the group $G$, $V(x)\in G$, so that there are ${\rm dim}(G)$ fields.  We then count as equivalent fields that differ by an element of the the subgroup, so $V(x)\sim V(x)h(x)$, where $h(x)\in H$.  To implement this equivalence, we demand that the theory be gauge invariant under local $h(x)$ transformations $V(x)\rightarrow V(x)h(x)$.  There are ${\rm dim}(H)$ gauge transformations, so the number of physical Goldstone bosons will be ${\rm dim}(G)-{\rm dim}(H)={\rm dim}({ G/H})$, the expected number. 

The global $G$ transformations act on the left as $V(x)\rightarrow gV(x)$, where $g\in G$.  The theory should therefore be invariant under the symmetries
\be V(x)\longmapsto g V(x)h^{-1}(x),\ee
where $g$ is a global $G$ transformation, and $h^{-1}(x)$ (written as an inverse for later convenience) is a local $H$ transformation.

A Lie group, G, possesses a distinguished left-invariant Lie algebra-valued 1-form, the so-called Maurer--Cartan form, given by $V^{-1}\rd V$.  Since this is Lie algebra-valued we may expand over a basis $\{V_I,Z_a\}$ where $\{V_I\}$, $I=1,\ldots,{\rm dim}(H)$ is a basis of the Lie algebra $\mathfrak{h}$ of H, and $\{Z_a\}$, $a=1,\ldots,{\rm dim}({G/H})$ is any completion to a basis of $\mathfrak{g}$.  We expand the Maurer--Cartan form over this basis,
\be\label{maurercartanf}
V^{-1}\rd V = \omega_V^I V_I + \omega_Z^a Z_a \ ,
\ee
where $\omega_V^I$ and $\omega_Z^a$ are the coefficients, which depend on the fields and their derivatives.  The Maurer--Cartan form (\ref{maurercartanf}), and hence the coefficients in the expansion on the right hand side, are invariant under global $G$ transformations. 

Under the local $h(x)$ transformation, the pieces $\omega_V\equiv\omega_V^I V_I$ and $\omega_Z\equiv\omega_Z^I Z_I$  transform as
\begin{align}
\omega_Z &\longmapsto h\, \omega_Z h^{-1},\nonumber\\
\omega_V &\longmapsto h\, \omega_V h^{-1}+ h\, \rd h^{-1}~.
\end{align}
We see that $\omega_Z$ transforms covariantly as the adjoint representation of the subgroup, and we use it as the basic ingredient to construct invariant Lagrangians~\cite{Coleman:1969sm, Callan:1969sn, volkov, xthschool}.
On the other had, $\omega_V$ transforms as a gauge connection.\footnote{This is a reflection of the well-known fact that the pullback of the Maurer--Cartan form defines a natural $H$-connection on $G/H$ \cite{Camporesi:1990wm, de Azcarraga:1997gn, nomizu}.}  If we have additional matter fields $\psi(x)$ which transform under some linear representation $D$ of the local group $H$ (and do not change under global $G$ transformations), 
\be 
\psi \longrightarrow D\left(h\right)\psi \ ,
\ee
we may construct a covariant derivative using $\omega_V$ via
\be 
{\cal D}\psi\equiv\rd\psi+D(\omega_V)\psi,\ \ \ \ ~~~~~~~{\cal D}\psi\rightarrow D\left(h\right){\cal D}\psi \ .
\ee
Thus, the most general Lagrangian is any Lorentz and globally $H$-invariant scalar constructed from the components of $\omega_Z$, $\psi$, and the covariant derivative,
\be 
{\cal L}\left({\omega_{Z}}^I_\mu,\psi,{\cal D}_\mu\right) \ .
\ee

To obtain a theory with global $G$ symmetry, we fix the $h(x)$ gauge symmetry by imposing some canonical choice for $V(x)$, which we call $\tilde V(x)$.  This canonical choice should smoothly pick out one representative element from each coset, so $\tilde V(x)$ contains ${\rm dim}({G/H})$ fields.  In general, a global $g$ transformation will not preserve this choice, so a compensating $h$ transformation---depending on $g$ and $\tilde V$---will have to be made at the same time to restore the gauge choice.  The gauge fixed theory will then have the global symmetry
\be \tilde V(x)\longmapsto g\tilde V(x)h^{-1}(g,\tilde V(x)).\label{trans1}\ee

If we can choose the parametrization such that the transformation (\ref{trans1}) is linear in the fields $\tilde V$ only when $g\in H$, then we will have realized the symmetry breaking pattern $G\rightarrow H$.  When the commutation relations of the algebra are such that the commutator of a broken generator with a generator of $H$ is again a broken generator $[V_I,Z]\sim Z$, (which is true if $G$ is a compact group), one way to accomplish this is to choose the parametrization
\be
\tilde V(x) = e^{\xi(x)\cdot Z}~ \ .
\label{gsection}
\ee
Here the real scalar fields $\xi^a(x)$ are the $\dim(G/H)=\dim G-\dim H$ different Goldstone fields associated with the symmetry breaking pattern. Under left action by some $ g \in  G$, (\ref{trans1}) gives the transformation law for the $\xi^a(x)$ as,
\be
e^{\xi\cdot Z}\rightarrow e^{\xi'\cdot Z}=ge^{\xi\cdot Z}  h^{-1}(g,\xi) \ ,
\label{cosetransformation}
\ee
As can be seen using the Baker--Campbell--Hausdorff formula and the commutation condition $[V_I,Z]\sim Z$, the action on $\xi$ is linear when $g\in H$.

\subsection{Spontaneously broken space-time symmetries\label{spontbrokenspacetimesymmetries}}
In the preceding subsection we reviewed the case of spontaneously broken internal symmetries. Galileons, however, arise as Goldstone modes of spontaneously broken space-time symmetries (the non-linear symmetries (\ref{shiftsym}) do not commute with the Poincar\'e generators).  Consequently, we must extend the coset procedure to account for subtleties involved in non-linear realizations of  symmetries which do not commute with the Poincar\'e group. This was worked out comprehensively by Volkov \cite{volkov} and is reviewed nicely in \cite{xthschool}. While the construction is generally similar to the internal symmetry case, the main subtlety is that now we must explicitly keep track of the generators of space-time symmetries in the coset construction.

Following \cite{xthschool}, we assume that our full symmetry group $G$ contains the unbroken generators of space-time translations $P_\alpha$, unbroken Lorentz rotations $J_{\alpha\beta}$, an unbroken symmetry subgroup $H$ generated by $V_I$ (which all together form a subgroup), and finally the broken generators denoted by $Z_a$. The broken generators may in general be a mix of internal and space-time symmetry generators. As before, we want to parameterize the coset $G/H$, but the parameterization now takes the form \cite{volkov, xthschool,Low:2001bw}
\be
\tilde V = e^{x\cdot P}e^{\xi(x)\cdot Z} \ .
\label{spacetimecoset}
\ee
Note that we treat the translation generators on the same footing as the broken generators, with the coefficients simply the space-time coordinates.\footnote{This is little more than bookkeeping.  While the space-time translations $P_{\mu}$ are not spontaneously broken since their representation is linear on the fields $\xi$, the coordinates $x^{\mu}$ formally transform non-linearly under a translation $x^\mu\rightarrow x^\mu+\epsilon^\mu$ which merits their inclusion in the coset parameterization. One intuitive way to understand this is to think of Minkowski space as the coset Poincar\'e/Lorentz, as is pointed out in \cite{Ogievetsky:1973,Low:2001bw}.  For the remainder of the paper, translation generators $P_{\mu}$ whose ``Goldstone" is a coordinate $x^{\mu}$ will be referred to as ``unbroken," while the remaining translations $P_{A}$ will be referred to as ``broken". }  
As in the case of the internal symmetries, under left action by some $g \in$ G, (\ref{spacetimecoset}) transforms non-linearly
\be
e^{x\cdot P}e^{\xi(x)\cdot Z} \longmapsto e^{x'\cdot P}e^{\xi'(x')\cdot Z}=g\, e^{ x\cdot P}e^{\xi(x)\cdot Z} h^{-1}( g,\xi(x))~,
\label{spacetimegtransformation}
\ee
where $h(g,\xi(x))$ belongs to the unbroken group spanned by $V_I$ and $J_{\mu\nu}$, but has dependence on $\xi$.

As in the internal symmetry case, the object in which we are interested is the Maurer--Cartan form
\be\label{maurercartanst}
\tilde V^{-1}\rd \tilde V = \omega_P^{\alpha} P_\alpha + \omega_Z^a Z_a + \omega_V^I  V_I+{1\over 2}\omega_J^{\alpha\beta} J_{\alpha\beta}~,
\ee
where we have again expanded in the basis of the Lie algebra $\mathfrak g$.  We may act with the transformation (\ref{spacetimegtransformation}) to determine that the components, $ \omega_P \equiv  \omega_P^{\alpha} P_\alpha,\ \omega_Z\equiv\omega_Z^a Z_a,\  \omega_V\equiv \omega_V^I  V_I,\  \omega_J\equiv {1\over 2}\omega_J^{\alpha\beta} J_{\alpha\beta}$ of the Maurer--Cartan 1-form transform as \cite{xthschool}
\begin{align}
\nonumber
\omega_P&\rightarrow h~\omega_P~h^{-1},\nonumber\\
\omega_Z&\rightarrow h~\omega_Z~ h^{-1},\nonumber\\
\omega_V+\omega_J&\rightarrow h\left(\omega_V+\omega_J\right)h^{-1}+h\, \rd h^{-1}~.\label{MCtransformationsSpacetimecase}
\end{align}
The covariant transformation rule for $\omega_P$ and $\omega_Z$ tells us that these are the ingredients to use in constructing invariant Lagrangians \cite{volkov, xthschool, Low:2001bw}.  The form $\omega_P$, expanded in components is
\be \omega_P=\rd x^\nu {\(\omega_P\)}^{\ \alpha}_\nu P_\alpha,\label{vielbeinextract}\ee
Here the components ${\(\omega_P\)}^{\ \alpha}_\nu $ should be thought of as an invariant vielbein, with $\alpha$ a Lorentz index, from which we can construct an invariant metric
\be g_{\mu\nu}= {\(\omega_P\)}^{\ \alpha}_\mu  {\(\omega_P\)}^{\ \beta}_\nu \eta_{\alpha\beta},\ee
and an invariant measure
\be
-\frac{1}{4!}\epsilon_{\alpha\beta\gamma\delta} \omega_P^{\alpha}\wedge\omega_P^{\beta}\wedge\omega_P^{\gamma}\wedge\omega_P^{\delta} = \rd^4x\sqrt{-g} ~.
\label{measurec}
\ee
The form $\omega_Z$, expanded in components 
\be \omega_Z=\rd x^\mu  {\(\omega_Z\)}^{\ a}_\mu Z_a,\ee
yields the basic ingredient ${\mathcal D}_\alpha\xi^a$, the covariant derivative of the Goldstones, through
\be \label{basiccov} {\(\omega_Z\)}^{\ a}_\mu={\(\omega_P\)}^{\ \alpha}_\mu {\mathcal D}_\alpha \xi^a.\ee

We can construct covariant derivatives ${\cal D}$ for matter fields $\psi$, transforming as some combined Lorentz and $H$ representation, which we call $D$, by using $\omega_V+\omega_J$ as a connection,
\begin{align}
\omega_P^\alpha\mathcal D_\alpha\psi &= \rd\psi+D(\omega_V)\psi+D(\omega_J) \psi~.
\end{align}
This can also be used to take higher covariant derivatives of the Goldstones.
From these pieces, $e_{\mu}^{\ \alpha}$, ${\mathcal D}_\alpha\xi^a$, $\psi$ and ${\mathcal D}_\alpha$, we can build the most general invariant Lagrangian by combining them in a Lorentz and $H$ invariant way, and then multiplying against the invariant measure \eqref{measurec}.
\subsection{Inverse Higgs constraint}
There is another subtlety that arises in extending the coset construction to the case of space-time symmetries---there can be non-trivial relations between different Goldstone modes leading to fewer degrees of freedom than na\"ive counting would suggest. This is the well-known statement that the counting of massless degrees of freedom in Goldstone's theorem fails in the case of broken space-time symmetries \cite{volkov,Ivanov:1975zq,Nielsen:1975hm,Low:2001bw, Bellucci:2002ji, McArthur:2010zm,Watanabe:2012hr,Hidaka:2012ym}; that is, the number of Goldstone modes will not in general be equal to $\dim(G/H)$.  This phenomenon is sometimes referred to as the  {\it inverse Higgs effect} \cite{Ivanov:1975zq}.

Accounting for this is simple---if the commutator of an unbroken translation generator with a broken symmetry generator, say $Z_{1}$, contains a component along some linearly independent broken generator, say $Z_{2}$,
\be
\left[P, Z_{1}\right] \sim Z_{2}+\cdots~,
\ee
(where the dots represent a component along the broken directions),
 it is possible to eliminate the Goldstone field corresponding to the generator $Z_1$ \cite{Low:2001bw, Ivanov:1975zq, McArthur:2010zm}. The relation between the Goldstone modes is obtained by setting the coefficient of $Z_{2}$ in the Maurer--Cartan form to zero.

 This is a covariant constraint; {\it i.e.}, it is invariant under G because the Maurer--Cartan form itself is invariant (often, the inverse Higgs constraint is imposed automatically in a constructed Lagrangian because it is equivalent to integrating out the redundant Goldstone field via its equation of motion \cite{McArthur:2010zm}).  We will need to use the inverse Higgs constraint in constructing the galileons.

\section{Cohomology\label{cohomologysection}}
As we shall see, the galileon terms are in fact not captured by the coset construction of the previous section.  This is essentially due to the fact that the coset construction produces Lagrangians which are strictly invariant under the desired symmetries, but the galileon Lagrangians are not strictly invariant---they change by a total derivative (so the action is still invariant).  As we shall also see, it will turn out that they can be thought of and categorized as non-trivial elements of Lie algebra cohomology.

In this section, we introduce the necessary concepts and definitions of Lie algebra cohomology and relative Lie algebra cohomology needed for classifying the galileons.  For a more comprehensive introduction, including applications, see \cite{deAzcarraga:1998uy}. 
 
\subsection{Lie algebra cohomology}
Given a Lie algebra $\mathfrak g$, an $n$-co-chain, $n=0,1,2,\ldots$, is a totally anti-symmetric multi-linear mapping $\omega_n: {\bigwedge}^{n}\mathfrak{g}\to {\mathbb{R}}$, taking values in the reals.\footnote{In general, one can consider the case in which the co-chains take values in an arbitrary vector space on which acts a non-trivial representation of $\mathfrak{g}$, but we do not need that here.}  The space of $n$-co-chains is denoted $\Omega^n(\mathfrak g)$. One then forms a co-boundary operator $\delta_{n}:\Omega^{n}(\mathfrak g)\to\Omega^{n+1}(\mathfrak g)$ whose action is defined by \cite{deAzcarraga:1998uy}
\begin{align}
\delta\omega(X_{1},X_{2},\ldots,X_{n+1})&=\sum_{j,k=1 \atop j<k}^{n+1}(-1)^{j+k}\omega([X_{j},X_{k}],X_{1},\ldots,\hat{X}_{j},\ldots,\hat{X}_{k},\ldots,X_{n+1}),
\end{align}
for $X_{1},X_2,\ldots\in\mathfrak{g}$ and where $\hat{X}$ means the argument is omitted, and $[\ ,\ ]$ is the Lie algebra commutator.  The first few instances are
\bea
\nonumber
&&\delta\omega_0 (X_1)=0,\\
\nonumber
&& \delta \omega_1(X_1,X_2)=-\omega_1([X_1,X_2]),\\
\nonumber
&& \delta \omega_2(X_1,X_2,X_3)= -\omega_2([X_1,X_2],X_3)+ \omega_2([X_1,X_3],X_2) -\omega_2([X_2,X_3],X_1),\\
&&~~~~\vdots
\eea
One can show, using the Jacobi identity $[X_1,[X_2,X_3]]+[X_2,[X_3,X_1]]+[X_3,[X_1,X_2]]=0$, that the co-boundary operator is nilpotent
\be 
\delta^{2}=0 \ .
\ee
Thus we have ${\rm Im}_{\delta_{n-1}}\(\Omega^{n-1}\)\subset {\rm Ker}_{\delta_n}\(\Omega^{n}\)$, and we can define the cohomology spaces
\be 
H^n(\mathfrak g)={{\rm Ker}_{\delta_n}\(\Omega^{n}(\mathfrak g)\) \over {\rm Im}_{\delta_{n-1}}\(\Omega^{n-1}(\mathfrak g)\) }\  .
\ee

There is another way to represent the co-boundary operator that is often more convenient when we have an explicit basis.  Let $\{e_i\}$, $i=1,\cdots,{\rm dim}(\mathfrak g)$, be a basis for the Lie algebra $\mathfrak g$.  The structure constants $c_{ij}^{\ \ k}$ are given by 
\be 
[e_i,e_j]=c_{ij}^{\ \ k}e_k \ .
\ee
They are anti-symmetric in their first indices, $c^{\ \ k}_{ij}=-c^{\ \ k}_{ji}$.  The Jacobi identity becomes $c^{\ \ m}_{il}c^{\ \ l}_{jk}+c^{\ \ m}_{jl}c^{\ \ l}_{ki}+c^{\ \ m}_{kl}c^{\ \ l}_{ij}=0$.  Let $\{\omega^{i}\}$ be a basis of the dual space $\mathfrak{g}^{*}$, dual to the basis $\{e_i\}$, so that $\omega^i(e_j)=\delta^i_j$.  Then we can write any $n$-co-chain $\omega_{n}$ as sums of wedge products of the $\omega^{i}$,
\be 
\label{omegacomp} \omega_n={1\over n!} \Omega_{i_1i_2\cdots i_n}\omega^{i_1}\wedge\omega^{i_2}\wedge\cdots\wedge\omega^{i_n} \ ,
\ee
where $\Omega_{i_1i_2\cdots i_n}$ is the totally anti-symmetric tensor of coefficients. The action of the co-boundary operator on a single $\omega^{i}$ is given by 
\be 
\delta \omega^{i}=-\frac{1}{2}c_{jk}^{~~i}\omega^{j}\wedge\omega^k \ ,
\ee
and is extended to wedge products of multiple $\omega$'s by using linearity and the Leibniz product rule, where we are careful to include the addition of a minus sign every time $\delta$ has to pass through an $\omega$.\footnote{The co-boundary operator, $\delta$, is an anti-derivation on the algebra of co-chains.}  For example, we have $\delta\left(\omega^i\wedge\omega^j\right)=-\frac{1}{2}c_{kl}^{~~i}\omega^{k}\wedge\omega^l\wedge\omega^j+\frac{1}{2}c_{kl}^{~~j}\omega^{i}\wedge\omega^k\wedge\omega^l.$  In terms of components, we have
\be \(\delta \Omega\)_{i_1\cdots i_{n+1}}=-{n(n+1)\over 2}c_{[i_1i_2}^{\ \ \ \ j}\Omega_{|j|i_3\cdots i_{n+1}]}.\ee

Lie algebra cohomology also has a geometric interpretation.\footnote{In this geometric context, Lie algebra cohomology is known as Chevalley--Eilenberg Cohomology \cite{Chevalley:1948zz}.}  Consider the simply connected Lie group $G$ associated to the Lie algebra $\mathfrak{g}$.  The space of $p$-forms on $G$ which are invariant under the left action of $G$ on itself can be identified with the co-chains of Lie algebra cohomology.  In fact, there is one left invariant 1-form for each generator of the Lie algebra, and wedging them together in all ways generates all the invariant $p$-forms.  The usual exterior derivative operator on $G$, $\rd_{p}: \Omega^{p}(G)\to  \Omega^{p+1}(G)$ satisfies ${\rm d}\omega^{i}=-\frac{1}{2}c_{jk}{}^{i}\omega^{j}\wedge\omega^{k}$,
and can be identified with the operator $\delta$ of Lie algebra cohomology.  Thus, Lie algebra cohomology counts the number of left-invariant forms on $G$ which cannot be written as the exterior derivative of a form which is also left-invariant.

\subsection{Relative Lie algebra cohomology}

For characterizing symmetry breaking to a subalgebra, we will need a slightly more refined notion of Lie algebra cohomology, known as relative Lie algebra cohomology.  Consider a subalgebra $\mathfrak{h}\subset\mathfrak{g}$.  We define the space of relative co-chains $\Omega^n(\mathfrak{g},\mathfrak{h})$, as the subspace of co-chains satisfying the following two conditions,
\begin{align}\label{relativcondition1} 
&\Omega_{n}(V,X_{2},\ldots,X_{n})=0 \ ,  \\
&\Omega_{n}([V,X_{1}],X_{2},\ldots,X_{n})+\Omega_{n}(X_{1},[V,X_{2}],\ldots,X_{n})+\cdots+\Omega_{n}(X_{1},X_{2},\ldots,[V,X_{n}])=0 \ , \nonumber \\ 
&\ \ \ {\rm for\ all}\ \ \,V\in\mathfrak{h},\ {\rm and}\    \ X_2,\cdots,X_n\in\mathfrak{g}\ .\label{invarianceundersubgroup}
\end{align}
The first requirement says that if any of the arguments lie completely in $\mathfrak{h}$, then we get zero.  This means that the form is well defined on the quotient $\mathfrak{g}/\mathfrak{h}$.  
Equivalently, the $n$-co-chains are only constructed from wedging together one-forms which annihilate $\mathfrak{h}$.  To see what this means in terms of components, choose a basis $\{h_I,f_a\}$ for $\mathfrak{g}$, where $\{h_I\}$, $I=1,\ldots,{\rm dim}(\mathfrak{h})$ is a basis of $\mathfrak{h}$ and $\{f_a\}$, $a=1,\ldots,{\rm dim}(\mathfrak{g}/\mathfrak{h})$ completes to a basis of $\mathfrak{g}$.  Let the dual basis be $\{\eta^I,\omega^a\}$.  To satisfy (\ref{relativcondition1}), forms are constructed by wedging together only the forms $\omega^a$, so the components $\Omega_{i_1\cdots i_n}$ of (\ref{omegacomp}) are zero if any of the indices are in the $\mathfrak{h}$ directions.

The second condition, in terms of components (\ref{omegacomp}), reads $c_{Ii_1}^{\ \ j}\Omega_{ji_2\cdots i_n}+c_{Ii_2}^{\ \ j}\Omega_{i_1j\cdots i_n}+\cdots+c_{Ii_n}^{\ \ j}\Omega_{i_1i_2\cdots j}=0.$
The combination of the two conditions (\ref{relativcondition1}) and (\ref{invarianceundersubgroup}) on the components, along with the fact that $c_{IJ}^{\ \ a}=0$ since $\mathfrak{h}$ is a subgroup, gives our final conditions in terms of components for a co-chain to be a relative co-chain,
\bea 
&&\Omega_{I i_2\cdots i_n}=0 \ , \\  
&&c_{Ia_1}^{\ \ b}\Omega_{ba_2\cdots a_n}+c_{Ia_2}^{\ \ b}\Omega_{a_1b\cdots a_n}+\cdots+c_{Ia_n}^{\ \ b}\Omega_{a_1a_2\cdots b}=0 \ . 
\label{condition2comp}
\eea

Given our basis, the matrices
\be 
\phi(h_I)_a^{\ b}=-c_{Ia}^{\ \ b}
\ee
form a representation of the subalgebra $\mathfrak{h}$,
\be 
 \phi(h_I) \phi(h_J)- \phi(h_J) \phi(h_I)=c_{IJ}^{\ \ K}\phi(h_K) \ ,
\ee
as can be straightforwardly shown using the Jacobi identity, as well as the condition $c_{IJ}^{\ \ a}=0$ which follows from the fact that $\mathfrak{h}$ is a subalgebra.  Thus, the indices $a,b,\ldots$ of the space $\mathfrak{g}/\mathfrak{h}$ furnish a representation of the subgroup $\mathfrak{h}$, and the condition (\ref{condition2comp}) says that the co-chain coefficients must be invariant tensors under the action of $\mathfrak{h}$ in this space.

The $\delta$ operator preserves the two conditions (\ref{relativcondition1}) and (\ref{invarianceundersubgroup}), so $\delta_n\left(\Omega^n(\mathfrak{g},\mathfrak{h})\right) \subset \Omega^{n+1}(\mathfrak{g},\mathfrak{h})$.  Thus we may think of $\delta$ as acting on the spaces $\Omega^n(\mathfrak{g},\mathfrak{h})$.  The cohomology classes of this action are denoted by $H^{p}(\mathfrak{g},\mathfrak{h})$ and the construction is known as {\it relative Lie algebra cohomology} \cite{deAzcarraga:1998uy},
\be 
H^n(\mathfrak{g},\mathfrak{h})={{\rm Ker}_{\delta_n}\(\Omega^{n}(\mathfrak{g},\mathfrak{h})\) \over {\rm Im}_{\delta_{n-1}}\(\Omega^{n-1}(\mathfrak{g},\mathfrak{h})\) }\ .
\ee
Each non-trivial element of $H^{d+1}(\mathfrak{g}, \mathfrak{h})$ corresponds to a Wess--Zumino term for a $d$-dimensional space-time \cite{de Azcarraga:1997gn,deAzcarraga:1998uy}.

Relative Lie algebra cohomology also has a geometric interpretation.  Consider the connected Lie group $G$ and subgroup $H$, corresponding to the algebra $\frak{g}$ and subalgebra $\frak{h}$.  We can think of the group $G$ as a fiber bundle, consisting of spaces $H$ fibered over the base space $G/H$.  The group $G$ acts naturally on $G/H$ (which is a homogeneous space with isotropy subgroup $H$).  The relative co-chains can be thought of as left invariant form on $G$ which are projectable to $G/H$, {\it  i.e.}, can be written as a pullback through the projection $G\rightarrow G/H$ of a unique form on $G/H$.  Thus they can be identified with invariant forms on $G/H$.  The operator $\delta$ can be identified with the usual exterior derivative ${\rm d}$, so relative Lie algebra cohomology counts the number of left-invariant forms on $G/H$ which cannot be written as the exterior derivative of a form which is also left-invariant.

\section{The galileon algebra\label{galalgebra}}
Having briefly introduced the standard techniques of non-linear realizations and made our acquaintance with Lie algebra cohomology, we now move on to the problem of principal interest---the construction of galileons using this machinery. In order to do this, however, we must first describe the symmetry algebra which the galileons non-linearly realize.  We will call this algebra the \textit{galileon algebra}.

A theory of $N$ galileons, $\pi^I$, $I=1,\ldots,N$, in $d$ space-time dimensions has the usual Poincar\'e invariance $\mathfrak{iso}(d-1,1)$, of a relativistic field theory, under which all the galileons are scalars,
\bea
&&\delta_{P_\mu}\pi^I=-\partial_\mu\pi^I \ ,\nonumber\\
&&\delta_{J_{\mu\nu}}\pi^I=(x_\mu\partial_\nu-x_\nu\partial_\mu)\pi^I \ .
\eea
These satisfy the usual commutation relations
\bea
\nonumber
&&[{P_\mu},{P_\nu}]=0 \ ,\\ \nonumber
&&[{J_{\mu\nu}}, {P_\sigma}]=\eta_{\mu\sigma} {P_\nu}-\eta_{\nu\sigma}  {P_\mu} \ ,\\ 
&&[ {J_{\mu\nu}}, {J_{\sigma\rho}}]=\eta_{\mu\sigma} {J_{\nu\rho}}- \eta_{\nu\sigma} {J_{\mu\rho}}+\eta_{\nu\rho} {J_{\mu\sigma}}- \eta_{\mu\rho} {J_{\nu\sigma}} \ .
\eea
There is also a linearly realized internal $\mathfrak{so}(N)$ symmetry under which the $\pi^I$ rotate in the fundamental representation,
\be 
\delta_{J_{IJ}}\pi^K=(\delta_{I}^K\delta_{JL}-\delta_{J}^K\delta_{IL})\pi^L \ ,
\ee
satisfying
\be 
[{J_{IJ}}, {J_{KL}}]=\delta_{IK} {J_{JL}}- \delta_{JK} {J_{IL}}+\delta_{JL} {J_{IK}}- \delta_{IL} {J_{JK}} \ .
\ee

Finally, there are the non-linear shift symmetries\footnote{For an interpretation of the conserved charges associated with these symmetries, see \cite{Nicolis:2010se}.}, 
\bea \delta_{C^I}\pi^J=\delta^{IJ}~,~~~~~~~~~~~~~~~~~~~~~~~~~\delta_{B_{\ \mu}^I}\pi^J=x_\mu\delta^{IJ}.
\eea
These shift symmetries commute amongst themselves, but have the following non-trivial commutation relations with the linearly realized symmetries,
\be
\begin{array}{ll}
\left[{P_\mu},{B_{\ \nu}^I}\right]=\eta_{\mu\nu}{C^I}, & \left[{J_{IJ}}, {C^K}\right]=\delta_I^K{C_J}-\delta_J^K{C_I}, \\
 \left[{J_{\mu\nu}}, {B_{\ \sigma}^I}\right]=\eta_{\mu\sigma} {B_{\ \nu}^I}-\eta_{\nu\sigma}  {B_{\ \mu}^I}, & \left[{J_{IJ}}, {B^K_{\ \mu}}\right]=\delta_I^K{B_{ J\mu}}-\delta_J^K{B_{ I\mu}} \ .
\end{array}
\ee
We will call the algebra satisfying these commutation relations the \textit{galileon algebra} in $d$ space-time dimensions, co-dimension $N$, and denote it by
\be 
\mathfrak{Gal}((d-1)+1, N) \ , 
\ee
where the first argument indicates that there are $d-1$ space dimensions, and $1$ time dimension.  Correspondingly, we will denote the galileon group by ${\rm Gal}((d-1)+1, N)$.

Consider first the special case when $d=1$, {\it i.e.}, a $0+1$ dimensional space-time.  The algebra $\mathfrak{Gal}(0+1, N)$ is the algebra of galilean transformations,  the symmetries of a free non-relativistic point particle moving in $N$ dimensions.  The $0+1$ dimensional space-time is the particle world-line, and the $N$ co-dimensions are the dimensions in which the particle moves.  The case $N=1$ gives the symmetries of the single field galileons (\ref{galileon0}), and the case $N\geq 2$ gives the symmetries of the $\mathfrak{so}(N)$ symmetric multi-field galileons (\ref{multigalileon0}).

\subsection{Geometric interpretation of the galileon algebra}

The galileon algebras can readily be given a geometric interpretation.  Recall that the Poincar\'e transformations can be thought of as the algebra of infinitesimal transformations that preserve the metric tensor $\eta_{\mu\nu} = {\rm diag}\left(-1, 1, 1, \ldots, 1\right)$.  The galileon algebra $\mathfrak{Gal}((d-1)+1, N)$ is the algebra of infinitesimal transformations of $\mathbb{R}^{d+N}$ that preserves \textit{two} different tensors, one covariant and one contravariant, 
\bea 
&& f_{\mu\nu} = {\rm diag}(\underbrace{-1, 1,\ldots, 1}_{d~{\rm slots}},0,\ldots 0), \\
 &&\tilde f^{\mu\nu} = {\rm diag}(0, 0,\ldots, 0,\underbrace{1,\ldots 1}_{N~{\rm slots}}) \ .
\eea 
 The finite form of this transformation can be given most easily by grouping the coordinates $(x^\mu,y^I)$ of $\mathbb{R}^{d+N}$ into a column vector with the addition of a $1$ in the last slot, and then giving the transformation in matrix form as
\be 
\left(\begin{array}{c}y^I \\x^\mu \\1\end{array}\right)\longmapsto \left(\begin{array}{ccc}R^I_{\ J} & b^{ I}_{ \ \nu} & c^I \\ 0 & \Lambda^\mu_{~~\nu} & p^\mu \\0 & 0 & 1\end{array}\right)\left(\begin{array}{c}y^J \\x^\nu \\1\end{array}\right).
\ee
Here $R^I_{~J}$ is a rotation matrix, $\Lambda^\mu_{~~\nu}$ is a Lorentz transformation, and $b^{ I}_{\ \mu}$, $c^I$ and $p^\mu$ are any real numbers.

\subsection{The galileon algebra as a contraction}

Yet another way to think of the galileon algebras as a Wigner--\.In\"on\"u contraction \cite{Inonu:1953sp} of the $(d+N)$-dimensional Poincar\'e algebra along $N$ of the spatial directions.  Physically, we can think of the galileons as describing a co-dimension $N$ brane, where the speed of light has been sent to infinity in the directions transverse to the brane, but remains finite in the directions along the brane.

To see this, begin with the $(d+N)$ dimensional Poincar\'e algebra $\mathfrak{iso}(d-1+N,1)$, with non-zero commutators
\begin{align}
\nonumber
\left [J_{BC},P_{A}\right ]&=\eta_{BA}P_{C}-\eta_{CA}P_{B},\\
\left [J_{AB},J_{CD}\right ]&=\eta_{AC}J_{BD}-\eta_{BC}J_{AD}+\eta_{BD}J_{AC}-\eta_{AD}J_{BC} \ ,
\end{align}
where $A,B\cdots=0,1,2,\ldots,d+N-1$ and $\eta_{AB} = {\rm diag}(-1,1,1,\ldots,1)$.  Now break apart the indices, using Greek letters for the first $d$ directions and Latin letters for the $N$ co-dimension directions,
\bea
&&\left [J_{\nu \rho },P_{\mu }\right ] =\eta_{\nu \mu }P_{\rho } - \eta_{\rho \mu }P_{\nu },\nonumber \\ 
&&\left [J_{\mu \nu },J_{\rho \sigma }\right ] = \eta_{\mu \rho }J_{\nu \sigma }-\eta_{\nu \rho }J_{\mu\sigma }+\eta_{\nu\sigma }J_{\mu \rho }-\eta_{\mu\sigma }J_{\nu \rho } ,\nonumber\\
\nonumber\\
&&\left [J_{I J},J_{K L}\right ] = \delta_{I K}J_{J L}-\delta_{J K}J_{I L}+\delta_{J L}J_{I K}-\delta_{I L}J_{J K}, \nonumber \\
&&\left [J_{J K},P_I\right ] = \delta_{J I}P_{K}-\delta_{K I}P_{J},\nonumber \\ 
\nonumber\\
&&\left [P_{\mu },J_{I\nu  }\right ] = \eta_{\nu \mu }P_{I } ,\nonumber\\
&&\left [P_{I },J_{J \rho }\right ] = -\delta_{I J}P_{\rho },\nonumber\\
&&\left [J_{I\nu },J_{K \sigma}\right ] = \delta_{I K }J_{\nu \sigma }+\eta_{\nu\sigma}J_{I K},\nonumber \\
&&\left [J_{K L},J_{I\nu}\right ] = \delta_{I K}J_{L\nu}-\delta_{I L}J_{K\nu} ,\nonumber\\
&& \left[J_{\rho \sigma},J_{I\nu }\right]=\eta_{\rho \nu }J_{I\sigma }-\eta_{\sigma\nu }J_{I \rho} \ .
\eea
The contraction is performed by introducing a parameter, $v$, which will be sent to infinity and which is inserted into the algebra by changing co-dimensional entries of $\eta_{AB}$ to $v$, so that $\eta_{AB} \rightarrow {\rm diag}~(-1,1,\ldots,1,v,v,\ldots,v)$ and making the following re-scalings
\be
P_{I}\longrightarrow vC_{I}~,~~~~~~~~~~~~
J_{I\nu}\longrightarrow vB_{I\nu}~,~~~~~~~~~~~~
J_{IJ}\longrightarrow vJ_{IJ}~.
\ee
After sending $v\rightarrow\infty$, the surviving non-trivial commutation relations are
\bea
&&\left [J_{\nu \rho },P_{\mu }\right ] =\eta_{\nu \mu }P_{\rho } - \eta_{\rho \mu }P_{\nu } \ ,\nonumber\\ 
&&\left [J_{\mu \nu },J_{\rho \sigma }\right ] = \eta_{\mu \rho }J_{\nu \sigma }-\eta_{\nu \rho }J_{\mu\sigma }+\eta_{\nu\sigma }J_{\mu \rho }-\eta_{\mu\sigma }J_{\nu \rho } \ ,\nonumber\\
\nonumber\\
&&\left [J_{I J},J_{K L}\right ] = \delta_{I K}J_{J L}-\delta_{J K}J_{I L}+\delta_{J L}J_{I K}-\delta_{I L}J_{J K} \ ,\nonumber\\
&&\left [J_{J K},C_I\right ] = \delta_{J I}C_{K}-\delta_{K I}C_{J} \ ,\nonumber\\ 
\nonumber\\
&&\left [P_{\mu },B_{I\nu  }\right ] = \eta_{\nu \mu }C_{I } \ ,\nonumber\\
&&\left [J_{K L},B_{I\nu}\right ] = \delta_{I K}B_{L\nu}-\delta_{I L}B_{K\nu} \ ,\nonumber\\
&& \left[J_{\rho \sigma},B_{I\nu }\right]=\eta_{\rho \nu }B_{I\sigma }-\eta_{\sigma\nu }B_{I \rho} \ .
\eea
These are exactly the commutations relations of $\mathfrak{Gal}((d-1)+1,N)$.

\section{Non-relativistic point particle moving in one dimension}

We now proceed with the coset construction, first considering the simplest case of a galileon: the one-dimensional non-relativistic free point particle.  We can think of this as a $0+1$ dimensional brane probing a non-relativistic $1+1$ dimensional bulk.  The Wess--Zumino nature of the kinetic term was pointed out in \cite{Gauntlett:1990nk} and is elegantly treated using jet bundles in \cite{deAzcarraga:1995jw}. Here, instead, we will derive equivalent results from the coset perspective.

We denote the single degree of freedom as $q(t)$, where $t$ is the one and only space-time coordinate.  We want to construct Lagrangians which are invariant under the algebra $\mathfrak{Gal}(0+1,1)$, which is three dimensional and whose generators act on $q(t)$ as follows
\be
\delta_C q = 1,~~~~~~~~~~~~~~~~~~~~~
\delta_B q = -t,~~~~~~~~~~~~~~~~~~~~~
\delta_P q = -\dot q~.
\label{freeparticlesymm}
\ee
Here $\delta_C$ is the shift symmetry on the field, $\delta_B$ is the analogue of the ``galilean" shift symmetry (the galilean boost of the non-relativistic particle) and $\delta_P$ is time translation of the field.  The algebra has only a single non-zero commutator\footnote{In relation to the $d$-dimensional algebra, we are defining $P\equiv P_0$, $B\equiv B_0$.}
\be
\left[B, P\right] = C~.
\label{1dgalalgebra}
\ee
The only transformation among (\ref{freeparticlesymm}) which is linear is $\delta_P$, the rigid translations of the line, so the breaking pattern is
\be \mathfrak{Gal}(0+1,1)\longrightarrow \mathfrak{iso}(1).\ee

To construct the most general Lagrangian which realizes these symmetries (\ref{freeparticlesymm}), we employ the coset construction for space-time symmetries reviewed in Section (\ref{spontbrokenspacetimesymmetries}).  The parametrization of the coset \eqref{spacetimecoset} is given by
\be
\tilde V = e^{t P}e^{q C+\xi B}~,
\ee
where $q$ is the Goldstone field that will become the physical field associated with the shift symmetry, and $\xi$ is the Goldstone field associated with the galilean boost symmetry.  Since the momentum $P$ is to be included in the coset, there is no subgroup $H$ to be linearly realized.  Thus the coset is the galilean group itself,
\be  
{\rm Gal}(0+1,1) \ .
\ee

Next we compute the Maurer--Cartan form (\ref{maurercartanst}),
\be
\omega = \tilde V^{-1}\rd \tilde V = \rd t P+\left(\rd q-\xi\rd t\right) C + \rd\xi B~,
\ee
and the component 1-forms used to build Lagrangians can then be read off as
\be
\label{mc1formfreeparticle}
\omega_P = \rd t~,~~~~~~~~~~~~~~~~~
\omega_C = \rd q-\xi\rd t~,~~~~~~~~~~~~~~~~~
\omega_B = \rd\xi~.
\ee
Now, it is important to note that there is an inverse Higgs constraint. Inspection of the only commutator of the algebra (\ref{1dgalalgebra}) shows that we can eliminate the $\xi$ field in favor of $q$ by setting $\omega_C=0$, implying the relation
\be
\xi = \dot q~.
\label{freeparticleinversehiggs}
\ee
Substitution into (\ref{mc1formfreeparticle}) then provides simplified expressions for the basis 1-forms
\be \omega_P = \rd t~,~~~~~~~~~~~~~~~~~~~~~~~~~
\omega_B = \ddot q\, \rd t~.
\ee
Thus, all the ingredients available for constructing invariant Lagrangians involve at least two derivatives on each $q$.  There is also the covariant derivative, but this turns out to be just ${d\over dt}$, so taking higher covariant derivatives will only add more time derivatives.  Lagrangians constructed in this way are all strictly invariant under the shift symmetries $\delta_B$ and $\delta_C$. 

This presents a puzzle, since we know that the free particle kinetic term, ${\cal L}={1\over 2} \dot q^2$, is also galilean invariant.  Although it is not invariant under $\delta_B$, it is invariant up to a total derivative, so it represents a perfectly good Lagrangian which is missed by the coset construction since it contains fewer than two derivatives per $q$.   Another missed example is the tadpole term ${\cal L}=q$, which changes up to a total derivative under both $\delta_B$ and $\delta_C$.  How do we construct these missing terms?

The answer is that these terms will appear as particular shift and boost invariant 2-forms which are themselves constructible from the Maurer--Cartan form (\ref{mc1formfreeparticle}).  These terms will live on the coset space, that is, the space in which $q$ and $\xi$ are considered as new coordinates in addition to the $t$ direction of space-time.  These 2-forms will also be total derivatives in this higher dimensional space, writable as $\rd$ of a 1-form.  The Lagrangian will be obtained by integrating this 1-form on the 1 dimensional subspace where $q=q(t)$ and $\xi=\xi(t)$.  

The symmetries on this space in our case are generated by the vector fields \cite{deAzcarraga:1995jw}\footnote{Note that the Lie bracket of left-invariant vector fields is minus the commutator of the algebra.}
\be
C = \partial_q~,~~~~~~~~~~~~~~~~~~~
B = \partial_{\xi}+t\partial_q~,~~~~~~~~~~~~~~~~~~~
P = \partial_t~.
\label{freeparticlevectors}
\ee
The components of the Maurer--Cartan form (\ref{mc1formfreeparticle}), where we treat $q$ and $\xi$ as independent coordinates, are the (left) invariant 1-forms on the coset space parametrized by $\{q,\xi,t\}$; that is we have $\pounds_X \omega=0$ where $X$ is any of the vector fields (\ref{freeparticlevectors}) and $\omega$ is any of the forms (\ref{mc1formfreeparticle}). 

Consider the invariant 2-forms, which are all obtained by wedging together all combinations of the invariant one-forms (\ref{mc1formfreeparticle}).  There are three of these, with the first being 
\be 
\omega_1^{\rm wz}=\omega_B\wedge\omega_C=\rd\xi\wedge\left(\rd q-\xi \rd t\right)\ .
\ee
We note that this can be written as the exterior derivative of a 1-form,
\be 
\omega_1^{\rm wz}=\rd\beta_1^{\rm wz}~,~~~~~~~~~~~~~~~~~~~~~~~~~\beta_1^{\rm wz}=\xi \rd q-{1\over 2}\xi^2 \rd t \ .
\ee
This 1-form can be used to construct an invariant action by pulling back to the surface space-time manifold $M$, defined by $q=q(t)$, $\xi=\xi(t)$, and then integrating,
\be
S^{\rm wz}_1 = \int_M\ \beta_1^{\rm wz} = \int \rd t\ \xi \dot q-{1\over 2}\xi^2~.
\ee
Imposing the inverse Higgs constraint $\xi=\dot q$ (or, equivalently, integrating out $\xi$), we recover the well-known kinetic term for the non-relativistic free point particle which was missed in the coset construction,
\be
S^{\rm wz}_1 = \int_M\ \beta_{1}^{\rm wz} = \int \rd t\ {1\over 2}\dot q^2~.
\ee

The tadpole term may be constructed similarly from the two form 
\be
\omega^{\rm wz}_2= \omega_C\wedge\omega_P =\rd q\wedge \rd t=\rd \beta_2^{\rm wz} \ ,~~~~~~~~~~~~~~~  \beta_2^{\rm wz}=q\rd t \ .
\ee
\be
S^{\rm wz}_2 = \int_M\ \beta_{2}^{\rm wz} = \int \rd t\ q~.
\ee

The final possible invariant 2-form constructible from the invariant one forms (\ref{mc1formfreeparticle}) is $\omega^{\rm wz}_3=\omega_B\wedge\omega_P=\rd\xi\wedge \rd t=\rd(\xi\rd t)$.  This leads to an action which is a total derivative once the Higgs constraint is imposed, and so nothing new results.  (This illustrates that the dimension of the relevant cohomology groups may not in general count the number of galileons exactly, but will only put an upper bound on the possible number.)

 In all cases, the 2-form $\omega^{\rm wz}$ is closed since it can be written as $\rd$ of a one form $\beta^{\rm wz}$ (so that we may use it to construct an action).  
Furthermore, the 2-form $\omega^{\rm wz}$ is by construction (left) invariant under the vector fields that generate the symmetries we are interested in (\ref{freeparticlesymm}).  However, the 1-form $\beta^{\rm wz}$ is not invariant---it shifts by a total $\rd$ (as it must since $\omega^{\rm wz}$ is invariant, $\omega^{\rm wz}=\rd\beta^{\rm wz}$, and de Rham cohomology is trivial on all the spaces we're considering), but this still leaves the action invariant.

The interesting 2-forms are therefore those which are invariant under the action of the vector fields (\ref{freeparticlevectors}) but which {\it cannot} be written as the exterior derivative of a 1-form which is itself invariant~\cite{deAzcarraga:1995jw} (since otherwise the corresponding 1-form on the boundary would be strictly invariant and would have already been captured by the coset construction).  They can thus be identified with non-trivial elements of the Lie algebra cohomology
\be  H^2\left(\mathfrak{Gal}(0+1,1)\right).\ee

Lagrangians constructed in this manner are what we call Wess--Zumino terms.  For a $d$-dimensional space-time, they are terms that correspond to non-trivial $d+1$ co-cycles in the cohomology of $\rd$ acting on invariant vector fields on the coset space (we will review this more carefully in the next section) \cite{Chevalley:1948zz}.

\section{Non-relativistic point particle moving in higher dimensions}

Now that we have understood a familiar system as the simplest example of a galileon theory, we are ready to apply the same techniques to the next most complicated case. We consider the point particle in higher co-dimensions, where in addition to space-time transformations, the fields can also rotate into each other in field space.  This describes a non-relativistic particle moving in the plane ${\mathbb R}^N$.  

The fields $q^I$ now have an extra index, $I=1,\cdots,N$,  the shift symmetries and time translation symmetry act as
\be
\delta_{C_J} q^I = \delta_{J}^I~,~~~~~~~~~~~~~~~~~~~~
\delta_{B_J} q^I = -t\delta_{J}^I~,~~~~~~~~~~~~~~~~~~~~
\delta_P q^I = -\dot q^I~,
\ee
and there is now an internal $\mathfrak{so}(N)$ symmetry,
\be  
\delta_{J_{IJ}}q^K=(\delta_{I}^K\delta_{JL}-\delta_{J}^K\delta_{IL})q^L \ .
\ee
The non-trivial commutation relations are 
\bea
&&\left [B_{I  },P\right ] = C_{I } \ , \nonumber \\
&&\left [J_{J K},C_I\right ] = \delta_{J I}C_{K}-\delta_{K I}C_{J} \ , \nonumber\\ 
&&\left [J_{K L},B_{I}\right ] = \delta_{I K}B_{L}-\delta_{I L}B_{K} \ ,\nonumber\\
&&\left [J_{I J},J_{K L}\right ] = \delta_{I K}J_{J L}-\delta_{J K}J_{I L}+\delta_{J L}J_{I K}-\delta_{I L}J_{J K} \ , 
\eea
and the symmetry breaking pattern is
\be 
\mathfrak{Gal}(0+1,N)\longrightarrow \mathfrak{iso}(1)\oplus\mathfrak{so}(N) \ .
\ee
The coset we are interested in is then
\be
{\rm Gal}(0+1,N)/{\rm SO}(N)~,
\ee
which is parameterized by \eqref{spacetimecoset},
\be
\tilde V = e^{tP}e^{q^IC_I+\xi^IB_I}\ .
\ee
(Recall that the unbroken generators, in this case the internal rotation generators, are not included in the coset, but the unbroken translations are).  
The Maurer--Cartan form \eqref{maurercartanst} is nearly the same as in the point particle case (\ref{mc1formfreeparticle}), except that some of the components now carry an extra internal index
\be
\omega_P = \rd t~,~~~~~~~~~~~~~~~~~
\omega_C^I = \rd q^I-\xi^I\rd t~,~~~~~~~~~~~~~~~~~
\omega_B^I = \rd\xi^I~. \label{sononeforms}
\ee
Similarly, the inverse Higgs constraint is now given by
\be  
\xi^I = \dot q^I \ .
\ee
As before, the only invariant form left for constructing actions is $\ddot q^I dt$, so all coset constructible actions contain at least two derivatives per field.

To construct the Wess--Zumino terms, we again create 2-forms by wedging together the 1-forms (\ref{sononeforms}), but now we must be sure that the forms are $\mathfrak{so}(N)$-invariant so that they are well defined on the coset.  This means that the $\mathfrak{so}(N)$ indices in (\ref{sononeforms}) must be contracted using $\mathfrak{so}(N)$ invariant tensors, and the only such tensors are $\delta_{IJ}$ and $\epsilon_{I_1\cdots I_N}$.  These forms will therefore be identified with the relative Lie algebra cohomology
\be 
H^2\left(  \mathfrak{Gal}(0+1,N),\mathfrak{so}(N)\right) \ .
\ee
We construct the kinetic terms of the fields by considering
\be 
\omega_1^{\rm wz}=\delta_{IJ}\omega_B^I\wedge\omega_C^J=\delta_{IJ}\rd\xi^I\wedge\left(\rd q^J-\xi^J \rd t\right) \ ,
\ee
which can be written as the exterior derivative of a 1-form,
\be 
\omega_1^{\rm wz}=\rd\beta_1^{\rm wz}~,~~~~~~~~~~~~~~~~~\beta_1^{\rm wz}=\delta_{IJ}\left(\xi^I \rd q^J-{1\over 2}\xi^I\xi^J\rd t\right) \ .
\ee
Pulling back to the surface space-time manifold $M$, defined by $q^I=q^I(t)$, $\xi^I=\xi^I(t)$, and then integrating, we have
\be
S_1^{\rm wz} = \int_M\ \beta_1^{\rm wz} = \int \rd t\ \delta_{IJ}\left(\xi^I \dot q^J-{1\over 2}\xi^I\xi^J\right)~,
\ee
and then imposing the inverse Higgs constraint $\xi^I=\dot q^I$ (or equivalently, integrating out $\xi^I$), we recover
\be
S^{\rm wz}_1 = \int_M\ \beta_1^{\rm wz} = \int \rd t\ {1\over 2}\delta_{IJ}\dot q^I\dot q^J~.
\ee

For $N\geq 2$ there is no longer a tadpole term, since the Lagrangian must be invariant under an ${\rm SO}(N)$ rotation of the fields $q^{I}$.  There are also no more non-trivial Wess--Zumino terms beyond the kinetic term (once the inverse Higgs constraints are imposed), with one exception:  for $N=2$ a novel Lagrangian appears involving the $\epsilon_{IJ}$ tensor,
\be
\label{2dexception} 
\omega^{\rm wz}=\epsilon_{IJ}\omega_B^I\wedge\omega_B^J =\epsilon_{IJ}\rd\xi^I\wedge \rd\xi^J=\rd\beta_2^{\rm wz},~~~~~~~~\beta_2^{\rm wz}=\epsilon_{IJ}\xi^I \rd\xi^J \ ,
\ee
\be 
S_2 = \int_M \beta^{\rm wz}_2=  \int \rd t\  \epsilon_{IJ}\xi^I \dot\xi^J \ ,
\ee
which upon imposing the inverse Higgs constraint becomes 
\be 
S_2 = \int_M \beta^{\rm wz}_2= \int \rd t\   \epsilon_{IJ}\dot q^I \ddot q^J \ .
\ee
Note that this is an example in which the imposition of the inverse Higgs constraint is not equivalent to integrating out redundant fields from the action.

Thus, in the bi-galileon case there is the extra Lagrangian $\mathcal L = \epsilon_{IJ}\dot q^J\ddot q^J$ which has third-order equations of motion, unlike the other galileons, so the relation between second order equations of motion and Wess--Zumino terms is not a perfectly tight one, though it holds in all other cases.  Even so, this term still describes fewer degrees of freedom (there are two fields each with third order equations, indicating six phase space degrees of freedom, or three real degrees of freedom) than the non-galileon terms, though it describes more than the kinetic term.

\section{Galileons}
\label{1fieldgalcoset}

We now perform the coset construction for galileons in four dimensions. This is the situation of greatest physical interest. We will consider the case which is inspired by a co-dimension 1 braneworld model, in which the galileons are related to the brane-bending mode into the bulk.

The galileons non-linearly realize the shift symmetries
\bea \delta_{C}\pi=1~,~~~~~~~~~~~~~~~~~~~~~~~~~\delta_{B^{\mu}}\pi=x^\mu \ ,
\eea
and we have the the non-trivial commutators
\bea
\left [P_{\mu },B_{\nu  }\right ] = \eta_{\mu \nu }C \ ,\ \ \ \ ~~~~~~~ \left[J_{\rho \sigma},B_{\nu }\right]=\eta_{\rho \nu }B_{\sigma }-\eta_{\sigma\nu }B_{ \rho} \ ,
\eea
which, along with the commutators of Poincar\'e transformations, fill out the galileon algebra $\mathfrak{Gal}(3+1,1)$.
The 4$d$ galileons non-linearly realize the symmetry breaking pattern
\be
\mathfrak{Gal}(3+1,1) \longrightarrow \mathfrak{iso}(3,1)~,
\ee
and the coset is parameterized by \eqref{spacetimecoset}
\be
\tilde V = e^{x\cdot P}e^{\pi C+\xi\cdot B}~.
\ee
Note that the linearly realized generators consists of only the Lorentz transformations, so we are working with the coset
\be \label{cosetgal01}
{\rm Gal}(3+1,1)/{\rm SO}(3,1) \ .
\ee
The coefficients of the components of the Maurer--Cartan form \eqref{maurercartanst} are
\be
\omega_P^\mu = \rd x^\mu~,~~~~~~~~~~~~
\omega_C = \rd\pi+\xi_\mu\rd x^\mu~,~~~~~~~~~~~~
\omega_B^\mu = \rd\xi^\mu~,~~~~~~~~~~~~
\omega_J^{\mu\nu} = 0~.~~~~~~~~~~~~
\label{singlefieldMCform}
\ee

As is the norm when breaking space-time symmetries, there are fewer Goldstone modes than na\"ive counting would lead us to believe. We have broken generators, $V_\mu$ and $C$, but we only have a single Goldstone mode $\pi$, and this can be seen from the presence of an inverse Higgs constraint---the commutator $\left[P_\mu, B_\nu\right] = \eta_{\mu\nu} C$ tells us that we may eliminate the $\xi_\mu$ field in favor of $\pi$ by setting $\omega_C = 0$, which leads to the relation
\be
\xi_\mu = -\partial_\mu\pi \ .
\ee
This allows us to write the components of the Maurer--Cartan form as 
\begin{align}
\omega_P^\mu = \rd x^\mu ~,~~~~~~~~~~~~~~~~~~~~~~~~~\omega_B^\mu &= -\rd x^\nu\partial_\nu\partial^\mu\pi~.
\end{align}
Since we can only build Lagrangians by using these ingredients (along with the higher covariant derivatives on $\pi$, which in this case are the same as ordinary derivatives), the field $\pi$ will only ever appear with at least 2 derivatives per field.  Thus we can never obtain the galileons from this construction, since the galileon terms \eqref{galileon0} all have fewer than two derivatives per field (the galileon Lagrangians with $n$ $\pi$'s have $2n-2$ derivatives).

The fact that they cannot be built by the coset construction is suggestive of the fact that the $4d$ galileons are Wess--Zumino terms in the same sense as the free particle kinetic term---they are 4-form potentials for non-trivial 5-co-cycles in Lie algebra cohomology.  The construction proceeds similarly to the $1d$ case. 

We work on the coset space, the space in which $\pi$ and $\xi^\mu$ are considered as new coordinates in addition to the $x^\mu$ directions of space-time. The Lagrangian will be obtained by integrating a Wess--Zumino form on the subspace where $\pi=\pi(x)$ and $\xi^\mu=\xi^\mu(x)$. 
The symmetries on the coset space are generated by the vector fields
\be
C = \partial_\pi~,~~~~~~~~~~~~~~~~~~~
B_\mu = \partial_{\xi^\mu}-x_\mu\partial_\pi~,~~~~~~~~~~~~~~~~~~~
P_\mu = \partial_\mu~.
\label{singlefieldvectors}
\ee
The components of the Maurer--Cartan form (\ref{singlefieldMCform}), where we treat $\pi$ and $\xi^\mu$ as independent coordinates, are the (left) invariant 1-forms on the coset space parametrized by $\{\pi,\xi^\mu,x^\mu\}$, so that we have $\pounds_X \omega=0$ where $X$ is any of the vector fields (\ref{singlefieldvectors}) and $\omega$ is any of the forms (\ref{singlefieldMCform}).

 To construct the Wess--Zumino terms, we create invariant 5-forms by wedging together the 1-forms (\ref{singlefieldMCform}). However, we must ensure that the forms are invariant under the Lorentz transformations $\mathfrak{so}(3,1)$ so that they are well defined on the coset.  This means that the Lorentz indices in (\ref{singlefieldMCform}) must be contracted using Lorentz invariant tensors, and the only such tensors are $\eta_{\mu\nu}$ and $\epsilon_{\mu\nu\rho\sigma}$.  From the cohomology perspective, this means that the galileon terms are members of the relative Lie algebra cohomology group 
\be 
H^{5}\left(\mathfrak{Gal}(3+1, 1), \mathfrak{so}(3,1)\right) \ .
\ee

Start by considering the invariant 5-form
\be 
\omega_1^{\rm wz}=\epsilon_{\mu\nu\rho\sigma}~\omega_ C\wedge\omega_P^\mu\wedge\omega_P^\nu\wedge\omega_P^\rho\wedge\omega_P^\sigma=\epsilon_{\mu\nu\rho\sigma} \rd\pi\wedge \rd x^\mu\wedge \rd x^\nu\wedge\rd x^\rho\wedge \rd x^\sigma \ ,
\ee
which can be written as the exterior derivative of a 4-form,
\be 
\omega_1^{\rm wz}=\rd\beta^{\rm wz}_1,\ \ \ \ \ ~~~~~~~~\beta^{\rm wz}_1= \epsilon_{\mu\nu\rho\sigma}\pi\rd x^\mu\wedge \rd x^\nu\wedge\rd x^\rho\wedge \rd x^\sigma \ .
\ee
Pulling back to the space-time manifold $M$, defined by $\pi=\pi(x)$, $\xi=\xi(x)$, and then integrating,
\be
S^{\rm wz}_1 = \int_M\ \beta^{\rm wz} _1= \int_M\ ~\pi\epsilon_{\mu\nu\rho\sigma}\rd x^\mu\wedge \rd x^\nu\wedge\rd x^\rho\wedge \rd x^\sigma \sim \int \rd^4x~\pi ~,
\ee
we recover the tadpole term, which is the first galileon.  Just as in the free particle case, the tadpole term appears as a 4-form which shifts by a total derivative under the symmetries and whose exterior derivative is a strictly invariant 5-form.

Next consider
\be 
\omega_2^{\rm wz}=\epsilon_{\mu\nu\rho\sigma}~\omega_ C\wedge\omega_B^\mu\wedge\omega_P^\nu\wedge\omega_P^\rho\wedge\omega_P^\sigma=\epsilon_{\mu\nu\rho\sigma} \left(\rd\pi+\xi_\lambda\rd x^\lambda\right)\wedge \rd\xi^\mu \wedge \rd x^\nu\wedge\rd x^\rho\wedge \rd x^\sigma \ ,
\ee
which can be written as the exterior derivative of a 4-form,\footnote{In showing this, it is helpful to use the identity
\be 
\epsilon_{\mu\nu\rho\sigma}\xi_\lambda \rd\xi^\mu\wedge \rd x^\lambda\wedge  \rd x^\nu\wedge\rd x^\rho\wedge \rd x^\sigma ={1\over 4}\epsilon_{\mu\nu\rho\sigma}\xi_\lambda\rd\xi^\lambda  \wedge \rd x^\mu\wedge  \rd x^\nu\wedge\rd x^\rho\wedge \rd x^\sigma=3! \xi_\mu d\xi^\mu\wedge \rd x^0  \wedge \rd x^1\wedge  \rd x^2\wedge\rd x^3 \ .
\ee
}
\be 
\omega^{\rm wz}_2=\rd\beta^{\rm wz}_2,\ \ \ \ \ \beta^{\rm wz}_2=\epsilon_{\mu\nu\rho\sigma}\left(\pi\rd\xi^\mu-\frac{1}{8}\xi^2\rd x^\mu\right)\wedge \rd x^\nu\wedge\rd x^\rho\wedge \rd x^\sigma \ .
\ee
Pulling back to the space-time manifold $M$ and integrating, we obtain
\be
S^{\rm wz}_2 = \int_M\ \beta^{\rm wz}_2 = 3! \int \rd^4x\ \left(\pi\partial_\mu\xi^\mu-\frac{1}{2}\xi^2\right) \ .
\ee
Imposing the Higgs constraint $\xi_\mu=-\partial_\mu\pi$ (or equivalently, integrating out $\xi^\mu$), we recover the kinetic term, which is the second galileon,
\be
S^{\rm wz}_2 \sim  \int \rd^4x\ (\partial\pi)^2 \ .
\ee
The construction of $\mathcal L_3$ is similar.  We consider
\be 
\omega_3^{\rm wz}=\epsilon_{\mu\nu\rho\sigma}~\omega_ C\wedge\omega_B^\mu\wedge\omega_B^\nu\wedge\omega_P^\rho\wedge\omega_P^\sigma=\epsilon_{\mu\nu\rho\sigma} \left(\rd\pi+\xi_\lambda\rd x^\lambda\right)\wedge \rd\xi^\mu \wedge \rd \xi^\nu\wedge\rd x^\rho\wedge \rd x^\sigma \ ,
\ee
which can be written as the exterior derivative of a 4-form\footnote{In showing this, it is helpful to use the identity
\be -{2\over 3}\epsilon_{\mu\nu\rho\sigma}\xi_\lambda \rd\xi^\lambda\wedge \rd\xi^\mu \wedge \rd x^\nu \wedge \rd x^\rho\wedge \rd x^\sigma=\epsilon_{\mu\nu\rho\sigma}\xi_\lambda \rd\xi^\mu\wedge \rd\xi^\nu\wedge \rd x^\lambda\wedge \rd x^\rho\wedge \rd x^\sigma=-4\xi_\lambda \rd\xi^\lambda\wedge \rd\xi_\mu\wedge(\ast_4 \rd x^\mu),
\ee
where $\ast_4$ is the Hodge star on the space of $x^\mu$'s.
}
\be 
\omega^{\rm wz}_3=\rd\beta^{\rm wz}_3,\ \ \ \ \ \beta^{\rm wz}_3= \epsilon_{\mu\nu\rho\sigma}\left (\pi \rd \xi^{\mu}\wedge\rd \xi^{\nu}\wedge \rd x^{\rho}\wedge\rd x^{\sigma}-\frac{1}{3}\xi^{2}\rd\xi^{\mu}\wedge\rd x^{\nu}\wedge\rd x^{\rho}\wedge\rd x^{\sigma}\right )\ .
\ee
Pulling back to the space-time manifold $M$ and integrating yields
\be
S^{\rm wz}_3 = \int_M\ \beta^{\rm wz}_3 =  \int_M\ \rd^{4}x\, \Big[-2\pi\left[(\partial_{\mu}\xi^{\mu})^{2}-\partial_{\mu}\xi^{\nu}\partial_{\nu}\xi^{\mu }\right ]+2\xi_{\alpha}\xi^{\alpha}\partial_{\mu}\xi^{\mu } \Big ] \ .
\ee
Imposing the Higgs constraint $\xi_\mu=-\partial_\mu\pi$, and performing a $4d$ integration by parts, we recover the cubic galileon,
\be
S^{\rm wz}_3 \sim \int_M\ \rd^{4}x\, \square\pi(\partial\pi)^2 \ .
\ee
The pattern in now clear.  The expressions for $\mathcal L_4$ and $\mathcal L_5$ will be given by the forms 
\begin{align}
\omega_4^{\rm wz} &=\epsilon_{\mu\nu\rho\sigma}\omega_C\wedge\omega_B^\mu\wedge\omega_B^\nu\wedge\omega_B^\rho\wedge\omega_P^\sigma~,\nonumber\\
\omega_5^{\rm wz} &=  \epsilon_{\mu\nu\rho\sigma}\omega_C\wedge\omega_B^\mu\wedge\omega_B^\nu\wedge\omega_B^\rho\wedge\omega_B^\sigma~,
\end{align}
respectively.  From the cohomology perspective, the galileon terms are members of the relative Lie algebra cohomology group $H^{5}\left(\mathfrak{Gal}(3+1, 1), \mathfrak{so}(3,1)\right)$.

\subsection{$d$ dimensional galileons}

This procedure is easily generalized to $d$ space-time dimensions, in which case the breaking pattern is 
\be \mathfrak{Gal}((d-1)+1,1)\to\mathfrak{iso}(d-1,1),\ee
and the coset is 
\be {\rm Gal}((d-1)+1,1)/{\rm SO}(d-1,1).\ee 

The $n$-th single field galileon term descends from the $(d+1)$-form
\bea 
\omega_n^{\rm wz}&&=\epsilon_{\mu_1\cdots\mu_d}\omega_C\wedge\omega_B^{\mu_1}\wedge\cdots\wedge\omega_B^{\mu_{n-1}}\wedge\omega_P^{\mu_n}\wedge\cdots\wedge\omega_P^{\mu_d} \ , \nonumber\\
&&=\epsilon_{\mu_1\cdots\mu_d}\left (\rd\pi+\xi_{\lambda}\rd x^{\lambda}\right )\wedge\rd \xi^{\mu_1}\wedge\cdots\wedge\rd \xi^{\mu_{n-1}}\wedge\rd x^{\mu_n}\wedge\cdots\wedge\rd x^{\mu_d},
\eea
where the basis 1-forms are the $d$-dimensional versions of (\ref{singlefieldMCform}).  This is the total derivative of the non-invariant Wess--Zumino $d$-form\footnote{We use the identity,
\bea &&{1\over (d-n+2)!}\xi_\lambda d\xi^\lambda\wedge  \rd \xi^{\mu_1}\wedge\cdots\wedge\rd \xi^{\mu_{n-2}}\wedge\rd x^{\mu_{n-1}}\wedge\cdots\wedge\rd x^{\mu_d}\epsilon_{\mu_1\cdots\mu_d} \nonumber \\
&&=-{1\over (n-1)(d-n+1)!}\xi_\lambda dx^\lambda \wedge  \rd \xi^{\mu_1}\wedge\cdots\wedge\rd \xi^{\mu_{n-1}}\wedge\rd x^{\mu_{n}}\wedge\cdots\wedge\rd x^{\mu_d} \epsilon_{\mu_1\cdots\mu_d}  \nonumber \\
&&=\xi_\lambda d\xi^\lambda \wedge  \rd \xi_{\mu_1}\wedge\cdots\wedge\rd \xi_{\mu_{n-2}}\wedge \ast_d\left(\rd x^{\mu_1}\wedge\cdots\wedge\rd x^{\mu_{n-2}}\right),
\eea
where $\ast_d$ is the Hodge star on the space of $x^\mu$'s.
} (in the following expressions $n\geq 2$, the tadpole is easily treated as before),
\bea \omega_n^{\rm wz}&=&\rd \beta_n^{\rm wz}, \nonumber \\
\beta_n^{\rm wz} &=& \epsilon_{\mu_1\cdots\mu_d}\bigg(\pi\rd \xi^{\mu_1}\wedge\cdots\wedge\rd \xi^{\mu_{n-1}}\wedge\rd x^{\mu_n}\wedge\cdots\wedge\rd x^{\mu_d}\nonumber \\ 
&&-{(n-1)\over 2(d-n+2)}\xi^2 \rd \xi^{\mu_1}\wedge\cdots\wedge\rd \xi^{\mu_{n-2}}\wedge\rd x^{\mu_{n-1}}\wedge\cdots\wedge\rd x^{\mu_d}\bigg).
\eea

Pulling back to the space-time manifold $M$ and integrating yields
\bea
S^{\rm wz}_n = \int_M\ \beta^{\rm wz}_n =  \int_M\ \rd^{d}x\, && (d-n+1)!(n-1)!\pi \delta_{\mu_1}^{[\nu_1}\cdots\delta_{\mu_{n-1}}^{\nu_{n-1}]} \partial_{\nu_1}\xi^{\mu_1}\cdots \partial_{\nu_{n-1}}\xi^{\mu_{n-1}} \ \nonumber \\
&& -{n-1\over 2}(d-n+1)!(n-2)!\xi^2\delta_{\mu_1}^{[\nu_1}\cdots\delta_{\mu_{n-2}}^{\nu_{n-2}]} \partial_{\nu_1}\xi^{\mu_1}\cdots \partial_{\nu_{n-2}}\xi^{\mu_{n-2}}.  \nonumber \\
\eea
Imposing the Higgs constraint $\xi_\mu=-\partial_\mu\pi$, and integrating the last term by parts, we recover the general galileon (\ref{galileon0}),
\be
S^{\rm wz}_n \sim \int_M\ \rd^{d}x\, \ \pi \delta_{\mu_1}^{[\nu_1}\cdots\delta_{\mu_{n-1}}^{\nu_{n-1}]} \partial_{\nu_1}\partial^{\mu_1}\pi\cdots \partial_{\nu_{n-1}}\partial^{\mu_{n-1}}\pi. 
\ee

The $d$ dimensional galileon terms are members of the relative Lie algebra cohomology group 
\be 
H^{d+1}\left(\mathfrak{Gal}((d-1)+1, 1), \mathfrak{so}(d-1,1)\right) \ .
\ee

\section{Multi-galileons}

It is straightforward to extend the analysis to the multi-galileon case.  The action and commutation relations are those of the algebra $\mathfrak{Gal}(3+1,N)$ described in Section \ref{galalgebra}, and the galileons realize the symmetry breaking pattern 
\be \mathfrak{Gal}(3+1,N) \longrightarrow \mathfrak{iso}(3,1)\oplus \mathfrak{so}(N)~.\ee
The coset is parameterized by \eqref{spacetimecoset}
\be
\tilde V = e^{x\cdot P}e^{\pi^I\cdot C_I+\xi^I\cdot B_I}~,
\ee
and the linearly realized subgroup consists of the Lorentz transformations and the $\mathfrak{so}(N)$ rotations, so we are working with the coset
\be 
{\rm Gal}(3+1,N)/\left({\rm SO}(3,1)\oplus {\rm SO}(N)\right) \ .
\ee
The coefficients of the components of the Maurer--Cartan form \eqref{maurercartanst} are
\be
\omega_P^\mu = \rd x^\mu~,~~~~~~~~~~~~
\omega_C^I = \rd\pi^I+\xi^I_{\ \mu}\rd x^\mu~,~~~~~~~~~~~~
\omega_B^{I\mu} = \rd\xi^{I \mu}~,~~~~~~~~~~~~
\omega_J^{\mu\nu} = \omega_J^{IJ}=0~,~~~~~~~~~~~~
\label{multfieldfieldMCform}
\ee
and the inverse Higgs constraint is
\be
\xi^I_{\ \mu} = -\partial_\mu\pi^I \ ,
\ee
so that we again find that we cannot construct any terms with fewer than two derivatives per $\pi^I$. 

To construct the Wess--Zumino terms, we create invariant 5-forms by wedging together the 1-forms (\ref{multfieldfieldMCform}), making sure that the forms are invariant under both the Lorentz transformations $\mathfrak{so}(3,1)$ and the internal $\mathfrak{so}(N)$ transformations so that they are well defined on the coset.  The two possible 5-forms that lead to non-trivial Lagrangians for $N\geq 2$ are
\bea 
&&  \omega_2^{\rm wz}=\delta_{IJ}\epsilon_{\mu\nu\rho\sigma}~\omega_ C^I\wedge\omega_B^{J\mu}\wedge\omega_P^\nu\wedge\omega_P^\rho\wedge\omega_P^\sigma \ ,\nonumber \\
 && \omega_4^{\rm wz}=\delta_{IJ}\delta_{KL}\epsilon_{\mu\nu\rho\sigma}~\omega_ C^I\wedge\omega_B^{J\mu}\wedge\omega_B^{K\nu}\wedge\omega_B^{L\rho}\wedge\omega_P^\sigma \ ,
\eea
leading to the kinetic term, and the quartic term studied in \cite{Hinterbichler:2010xn}. 

From the cohomology perspective, the multi-galileon terms are members of the relative Lie algebra cohomology group 
\be 
H^5\left(\mathfrak{Gal}(3+1, N), \mathfrak{so}(3,1)\oplus  \mathfrak{so}(N)\right) \ .
\ee

Generalizing to $d$-dimensions, there are $d/2$ possible terms for $d$ even, and $(d+1)/2$ possible terms for $d$ odd.  The Wess--Zumino $(d+1)$-forms are
\be 
\omega_{2n}^{\rm wz}=\delta_{I_1J_1}\cdots\delta_{I_nJ_n}\epsilon_{\mu_1\cdots \mu_d}~\omega_ C^{I_1}\wedge\omega_B^{J_1\mu_1}\wedge\cdots\wedge\omega_B^{I_n\mu_{2n-2}}\wedge\omega_B^{J_n\mu_{2n-1}}\wedge\omega_P^{\mu_{2n}}\wedge\cdots\wedge\omega_P^{\mu_d} \ ,
\ee
which lead to the Lagrangian (\ref{multigalileon0}).  They are members of the relative Lie algebra cohomology group 
\be 
H^{d+1}\left(\mathfrak{Gal}((d-1)+1, N), \mathfrak{so}(d-1,1)\oplus  \mathfrak{so}(N)\right) \ .
\ee
Note that using $\epsilon_{I_1\cdots I_N}$ to contract indices gives nothing new, leading only to Lagrangians which are total derivatives (with the exception of $d=1$, $N=1$ in (\ref{2dexception})). 

\section{Counting the galileons}
\label{4dgalcohomology}

While the construction of the four dimensional single field galileons makes it hard to imagine any other possible galileon invariant Lagrangians (and it has been shown by other methods that there aren't any \cite{Nicolis:2008in}), it is good to have a formal check that we have indeed found every possible Wess--Zumino term. After all, every Lagrangian that is compatible with the symmetries of the theory should be included when constructing an effective field theory, and so proper bookkeeping and accounting of terms is a worthwhile endeavor.

In order to verify that we have found all possible Wess--Zumino terms, we want to compute the relative Lie algebra cohomology $H^{5}\left(\mathfrak{Gal}(3+1, 1), \mathfrak{so}(3,1)\right)$.  
Noting that (\ref{singlefieldMCform}) is a basis for left-invariant forms, we determine the action of the exterior derivative, ${\rm d}$, on these forms
\be
{\rm d}\omega_P^\mu = 0~,~~~~~~~~~~~~~~~
{\rm d}\omega_B^\mu = 0~,~~~~~~~~~~~~~~~
{\rm d}\omega_C = \eta_{\mu\nu}\omega_B^\mu\wedge \omega_P^\nu \ .
\ee
To meet the requirement of ${\rm SO}(3,1)$ invariance, all Greek indices must be contracted with $\eta_{\mu\nu}$ or $\epsilon_{\mu\nu\rho\sigma}$.  Then, the ${\rm SO}(3,1)$ invariant 5-co-cycles can be explicitly constructed and are given by
\begin{align}
\label{non-trivial5forms}
\nonumber
\omega_1 &= \epsilon_{\mu\nu\rho\sigma}\omega_C\wedge \omega_P^\mu\wedge \omega_P^\nu\wedge \omega_P^\rho\wedge \omega_P^\sigma \ ,\\
\nonumber
\omega_2 &= \epsilon_{\mu\nu\rho\sigma}\omega_C\wedge \omega_P^\mu\wedge \omega_P^\nu\wedge \omega_P^\rho\wedge \omega_B^\sigma \ ,\\
\omega_3 &= \epsilon_{\mu\nu\rho\sigma}\omega_C\wedge \omega_P^\mu\wedge \omega_P^\nu\wedge \omega_B^\rho\wedge\omega_B^\sigma \ ,\\
\nonumber
\omega_4 &= \epsilon_{\mu\nu\rho\sigma}\omega_C\wedge \omega_P^\mu\wedge \omega_B^\nu\wedge \omega_B^\rho\wedge \omega_B^\sigma \ ,\\
\nonumber
\omega_5 &= \epsilon_{\mu\nu\rho\sigma}\omega_C\wedge \omega_B^\mu\wedge \omega_B^\nu\wedge \omega_B^\rho\wedge \omega_B^\sigma\ .
\end{align}
It is clear that each of these forms are closed, $\rd\omega = 0$.
 Furthermore, due to the presence of a factor of $\omega_C$ in each form, none of these are expressible as the exterior derivative of a 4-form.  In order to not vanish there must have been exactly one factor of $\omega_C$ in the 4-co-chain, but such a form is not Lorentz invariant; therefore all of the 5-co-cycles in (\ref{non-trivial5forms}) are non-trivial elements of $H^{5}(\mathfrak{gal}(1+3,1), \mathfrak{so}(3,1))$.  
 
 This provides a formal check that there only exist the five galileon Lagrangians and we have not missed any other Wess--Zumino terms in our construction.  Similar remarks apply to all other dimensions and co-dimensions. 

\section{Conformal galileons}
The conformal galileon is a higher derivative theory of a single scalar field, with second order equations of motion, and which non-linearly realizes the conformal group. The relevant Lagrangians were first constructed in Sec. 3.1 of \cite{Nicolis:2008in},
\bea  
{\cal L}_1&=&-{1\over 4}e^{4\pi} \ , \nonumber \\
{\cal L}_2&=&-\half e^{2\pi}(\partial\pi)^2 \ ,\nonumber \\
{\cal L}_3&=&\frac{1}{2}(\partial {\pi})^2 \Box {\pi}+\frac{1}{4} (\partial {\pi})^4 \ ,\nonumber\\
{\cal L}_4&=&- \frac{1}{2}e^{-2 \pi}(\partial  \pi)^2
\([ \Pi]^2-[ \Pi^2]+{2\over 5}(-(\partial  \pi)^2 \Box  \pi+[ \pi^3])+{3\over 10}(\partial  \pi)^4\) \ , \nonumber \\
{\cal L}_5&=& -\half e^{-4 \pi} (\partial  \pi)^2
\Big[-[ \Pi]^3+3[ \Pi][ \Pi^2]-2[ \Pi^3]+3(\partial  \pi)^2([ \Pi]^2-[ \Pi^2])\nonumber\\
&&+\frac{30}{7}(\partial  \pi)^2(-(\partial  \pi)^2[ \Pi]+[ \pi^3])-\frac{3}{28}(\partial  \pi)^6\Big] \ .\label{DBIgalderivexpan}
\eea
We have used the notation $\Pi$ for the matrix of partials $\Pi_{\mu\nu}\equiv\partial_{\mu}\partial_\nu\pi$, and brakets denote traces, $[\Pi^n]\equiv Tr(\Pi^n)$, e.g. $[\Pi]=\square\pi$, $[\Pi^2]=\partial_\mu\partial_\nu\pi\partial^\mu\partial^\nu\pi$.  We've also defined $[\pi^n]\equiv \partial\pi\cdot\Pi^{n-2}\cdot\partial\pi$, e.g. $[\pi^2]=\partial_\mu\pi\partial^\mu\pi$, $[\pi^3]=\partial_\mu\pi\partial^\mu\partial^\nu\pi\partial_\nu\pi$.  Indices are raised and lowered with $\eta_{\mu\nu}$.

The conformal galileons linearly realize Poincar\'e symmetry,
\begin{align}
\label{confpoincare}
\nonumber
\delta_{P_\mu}\pi &= -\partial_\mu\pi~,\\
\delta_{J_{\mu\nu}}\pi &=  (x_\mu\partial_\nu - x_\nu\partial_\mu)\pi~,
\end{align}
while the conformal symmetry is non-linearly realized
\begin{align}
\label{confconf}
\nonumber
\delta_D\pi &= -1-x^\mu\partial_\mu\pi~,\\
\delta_{K_\mu}\pi &=-2x_\mu -(2x_\mu x^\nu\partial_\nu-x^2\partial_\mu)\pi~.
\end{align}
Taken together, the transformations satisfy the commutators of the conformal algebra $\frak{so}(4,2)$,
\begin{equation}
\begin{array}{ll}
\left[P_\mu, D\right] = P_\mu~,& \left[D, K_\mu\right] = K_\mu~,\\
\left[J_{\mu\nu}, K_\sigma\right] = \eta_{\mu\sigma}K_\nu-\eta_{\nu\sigma}K_\mu~, & \left[J_{\mu\nu},P_\sigma\right] =\eta_{\mu\sigma}P_\nu-\eta_{\nu\sigma}P_\mu~,\\
\left[K_\mu,P_\nu\right] = 2J_{\mu\nu}-2\eta_{\mu\nu}D~, & \left[J_{\mu\nu}, J_{\rho\sigma}\right] = \eta_{\mu\rho}J_{\nu\sigma}-\eta_{\nu\rho}J_{\mu\sigma}+\eta_{\nu\sigma}J_{\mu\rho}-\eta_{\mu\sigma}J_{\nu\rho}~. \label{conformalgebra}
\end{array} 
\end{equation}
The conformal galileons may be interpreted as the Goldstone field associated with the symmetry breaking pattern
\be
\frak{so}(4,2) \longrightarrow \frak{iso}(3,1)~.
\label{conformaltopoincare}
\ee
 As we shall see, it is possible to obtain the conformal galileon terms via the coset construction, with the exception of the term quartic in derivatives, ${\cal L}_3$, which appears as a Wess--Zumino term.  
 
The coset space is 
\be {\rm SO}(4,2)/{\rm SO}(3,1),\ee
which we parametrize as\footnote{This differs slightly from our general expression \eqref{spacetimecoset} since we write a product of exponentials for the broken generators rather than the exponential of a sum.  This just amounts to a different choice of parametrization for the coset.}
\be
\tilde V=e^{x\cdot P}e^{\pi D}e^{\xi\cdot K}.\label{conformalcosetpar}
\ee
Calculating the  Maurer--Cartan form \eqref{maurercartanst},
\be
\omega = \tilde V^{-1}{\rm d}\tilde V = \omega_P^\mu P_\mu + \omega_D D+\omega_K^\mu K_\mu + {1\over 2}\omega_J^{\mu\nu}J_{\mu\nu}~,
\ee
the components are found to be \cite{Hinterbichler:2012mv,volkov, Bellucci:2002ji,McArthur:2010zm}
 \begin{align}
\nonumber
 \omega_{P}^{\mu}&=e^{\pi}\rd x^\mu,\\
\nonumber
 \omega_{D}&=\rd\pi+2e^{\pi}\xi_{\mu}\rd x^{\mu},\\
\nonumber
 \omega^{\mu}_{K}&=\rd \xi^{\mu}+\xi^{\mu}\rd \pi+e^{\pi}\left (2\xi^{\mu}\xi_{\nu}\rd x^{\nu}-\xi^{2}\rd x^{\mu}\right ),\\
 \omega^{\mu\nu}_{J}&=-4e^\pi\left (\xi^{\mu}\rd x^{\nu}-\xi^{\nu} \rd x^{\mu}\right )\ .
 \end{align}
where indices have been raised and lowered with $\eta_{\mu\nu}$. 

Due to the commutator $\left[K_\mu,P_\nu\right] = 2J_{\mu\nu}-2\eta_{\mu\nu}D$, there is an inverse Higgs constraint, $\omega_D = 0$ yielding the relation
\be
\xi_\mu = -\frac{1}{2}e^{-\pi}\partial_\mu\pi~.
\label{conformalinversehiggs}
\ee
Plugging back into the Maurer--Cartan form, we have

\be
\begin{array}{l}
\omega_P^\mu = e^\pi {\rm d}x^\mu,\\
\omega_K^\mu = e^{-\pi}\left({1\over 2}\partial_\nu\pi\partial^\mu\pi\rd x^\nu-{1\over 2}\partial_\nu\partial^\mu\pi \rd x^\nu-{1\over 4}(\partial\pi)^2 dx^\nu\right),\\
\omega_J^{\mu\nu} = 2\left(\partial^\mu\pi {\rm d} x^\nu-\partial^\nu\pi{\rm d} x^\mu\right)~.
\end{array}
\label{confmcforms}
\ee

The vielbein \eqref{vielbeinextract} can be extracted from $\omega_P$,
\be e_\nu^{\ \alpha} = e^\pi\delta_\nu^\alpha,\ee
giving the invariant metric 
\be\label{confinmet}
g_{\mu\nu} = e_\mu^{\ \alpha} e_\nu^{\ \beta}\eta_{\alpha\beta}= e^{2\pi}\eta_{\mu\nu}.
\ee
The invariant measure (\ref{measurec}) is 
\be \sqrt{-g}=e^{4\pi}.\ee
The derivative \eqref{basiccov} associated to $\xi^\beta$ (here another Lorentz index $\beta$ plays the role of the index $a$ in Section \ref{spontbrokenspacetimesymmetries}) is given by the expression $\(\omega_K\)_\mu^{\ \beta} = e_\mu^{\ \alpha}\mathcal D_\alpha\xi^\beta~$.
By contracting with the vielbein, we can instead work with the object $\mathcal D_\mu\xi_\nu\equiv e_\mu^{\ \alpha}\mathcal D_\alpha\xi^\beta e_\nu^{\ \gamma}\eta_{\beta\gamma}=\(\omega_K\)_\mu^{\ \beta} e_\nu^{\ \alpha}\eta_{\beta\alpha}$,
\be
\mathcal D_\mu\xi_\nu =\frac{1}{2}\partial_\nu\pi\partial_\mu\pi-\frac{1}{2}\partial_\nu\partial_\mu\pi-\frac{1}{4}(\partial \pi)^2\eta_{\mu\nu}~.
\label{dxia}
\ee
We construct invariant Lagrangians by using $D_\mu\xi_\nu$, contracting up indices with the metric \eqref{confinmet} and multiplying by the measure (\ref{measurec}).

 Another method is used in \cite{Nicolis:2008in}\footnote{There is also a method called tractor calculus, which is designed for constructing realizations of conformal symmetry \cite{thomas1,thomas2,gover1,eastwood,Gover:2008pt,Gover:2008sw,Bonezzi:2010jr}.}.  The conformal galileons are constructed by forming diffeomorphism scalars of the conformal metric $g_{\mu\nu}=e^{2\pi}\eta_{\mu\nu}$.  This method is in fact completely equivalent to the coset construction, because we have for the Ricci tensor
 \be
R_{\mu\nu}(g) = 2\partial_\mu\pi\partial_\nu\pi-2\partial_\mu\partial_\nu\pi-\square\pi\eta_{\mu\nu}-2(\partial\pi)^2\eta_{\mu\nu}~,
\ee
which can be expressed in terms of the covariant derivative \eqref{dxia}, 
\be
R_{\mu\nu}(g)=4\mathcal D_\mu\xi_\nu+2\mathcal D_\rho\xi^\rho g_{\mu\nu} ~.
\ee
The Ricci scalar for the conformal metric is $R[g]=12\mathcal D_\rho\xi^\rho$, and the Riemann tensor gives nothing beyond the Ricci tensor because the Weyl tensor vanishes for the conformally flat metric \eqref{confinmet}.  Furthermore, higher covariant derivatives $\mathcal D$ in the coset are equivalent to higher covariant derivatives $\nabla(g)$ with respect to the metric \eqref{confinmet}.  We therefore see that the invariant actions constructible by the coset method correspond to all possible diffeomorphism scalars constructed from the metric $g_{\mu\nu} = e^{2\pi}\eta_{\mu\nu}$, its curvature tensors and its covariant derivative.
 
The zero derivative term in \eqref{DBIgalderivexpan} comes from the volume element
\begin{align}
\mathcal{L}_{1}\sim  \sqrt{-g}=e^{4\pi}~,
\end{align}
while the kinetic term comes from the Ricci curvature, after an integration by parts
\begin{align}
\mathcal{L}_{2}\sim \sqrt{-g}R= 6 e^{2\pi}(\partial\pi)^{2}~.
\end{align}
The terms ${\cal L}_4$ and ${\cal L}_5$ are constructed from particular curvature invariants of order $R^3$ and $R^4$, respectively \cite{Nicolis:2008in}.

The term ${\cal L}_3$, however, presents a problem.  It should be constructible from curvature invariants of order $R^2$, but all three curvature invariants which are quadratic in the Ricci curvature give the same contribution after integration by parts (and in fact, only two could have been independent since the Gauss-Bonnet combination $R^2-4 R^{\mu\nu}R_{\mu\nu}+R^{\mu\nu\rho\sigma}R_{\mu\nu\rho\sigma}$ is a total derivative) { \cite{Nicolis:2008in}.
\begin{align}
\left .\begin{array}{c}
\sqrt{-g}R^{2}\\
\sqrt{-g}R^{\mu\nu}R_{\mu\nu}\\
\sqrt{-g}R^{\mu\nu\rho\sigma}R_{\mu\nu\rho\sigma}
\end{array}\right \}&\propto (\square\pi)^2 + (\partial\pi)^4+2\square\pi(\partial\pi)^2 ~,
\label{naiveconformalL3s}
\end{align}
which is not of the form ${\cal L}_3$ and gives rise to higher order equations of motion due to the $(\square\pi)^{2}$ term.  It would thus appear that it is impossible to create the conformal galileon ${\cal L}_3$ by the coset method. 

However, one can create a linearly independent invariant Lagrangian by using a trick, as is done in \cite{Nicolis:2008in}.  We go to $d$ dimensions,\footnote{The $d$-dimensional metric is $e^{2\pi}$ times the $d$-dimensional Minkowski metric.} and consider the following combination of curvature invariants, 
\begin{align}
\frac{\sqrt{-g}}{(d-4)}\bigg(&\frac{R_{\mu\nu}^2}{(d-1)}-\frac{R^2}{(d-1)^2}\bigg)=e^{(d-4)\pi}\left((\square\pi)^2+\frac{(d-2)(3d-4)}{2(d-1)}\square\pi(\partial\pi)^2+\frac{(d-2)^3}{2(d-1)}(\partial\pi)^4\right)~.
\end{align}
This combination is finite in the limit $d\to4$ and leads to the Lagrangian
\be
\mathcal L \sim \frac{3}{4}(\square\pi)^2 +(\partial\pi)^4+ 2\square\pi(\partial\pi)^2~.
\ee
This combination is linearly independent of (\ref{naiveconformalL3s}), and can be used to subtract off the offending $(\square\pi)^2$ term giving the cubic galileon
\begin{align}
 \mathcal{L}_{3}&\sim (\partial\pi)^{4}+2\square\pi(\partial\pi)^{2} \ .
 \end{align} 
The fact that we must do this dimensional continuation to construct $\mathcal L_3$ is a harbinger of the fact that this is a Wess--Zumino term. The fact that Wess--Zumino terms are not captured by the coset construction appears here through the fact that we have to move away from four dimensions.  In fact, it is easy to show that $\mathcal{L}_{3}$ changes by a total derivative under the non-linear symmetries while the remaining Lagrangians are strictly invariant (modulo the total derivative associated with changing the field coordinates), so we expect the necessity of a Wess--Zumino type construction for $\mathcal{L}_{3}$.

Starting with the conformal algebra \eqref{conformalgebra}, we wish to compute the relative Lie algebra cohomology 
\be H^{5}(\mathfrak{so}(4,2),\mathfrak{so}(3,1)),\ee
 in order to catalog the possible Wess--Zumino terms. Recall from Section \ref{cohomologysection} that the basis forms which are dual to the Lie algebra vectors are written with upper indices and the forms which annihilate the vector subspace spanned by $\mathfrak{so}(3,1)$  are $\left \{D,K^{\mu},P^{\mu}\right \}$.  These are used to create $n$-co-chains for computing the relative Lie algebra cohomology.  The co-boundary operator $\delta$ acts on the basis forms as
 \begin{align}
\nonumber
 \delta D&=2\eta_{\mu\nu}K^{\mu}\wedge P^{\nu}~,\\  
   \delta P^{\mu}&=D\wedge P^{\mu}+2P^{\beta}\wedge J^{\alpha\mu}\eta_{\alpha\beta}~,\\ 
\nonumber
    \delta K^{\mu}&=-D\wedge K^{\mu}+2K^{\beta}\wedge J^{\alpha\mu}\eta_{\alpha\beta}~.
 \end{align}
We can construct the following six $\mathfrak{so}(3,1)$ invariant 5-co-chains
 \begin{align}
\nonumber
 \omega_{1}&=\epsilon_{\mu\nu\rho\sigma}D\wedge P ^{\mu}\wedge  P ^{\nu}\wedge P ^{\rho}\wedge P ^{\sigma},\\
\nonumber
 \omega_{2}&=\epsilon_{\mu\nu\rho\sigma}D\wedge P ^{\mu}\wedge  P ^{\nu}\wedge P ^{\rho}\wedge K ^{\sigma},\\
\nonumber
 \omega_{3}&=\epsilon_{\mu\nu\rho\sigma}D\wedge P ^{\mu}\wedge  P ^{\nu}\wedge K ^{\rho}\wedge K ^{\sigma},\\
\nonumber
 \omega_{4}&=\epsilon_{\mu\nu\rho\sigma}D\wedge P ^{\mu}\wedge  K ^{\nu}\wedge K ^{\rho}\wedge K ^{\sigma},\\
\nonumber
 \omega_{5}&=\epsilon_{\mu\nu\rho\sigma}D\wedge K ^{\mu}\wedge  K ^{\nu}\wedge K ^{\rho}\wedge K ^{\sigma},\\
 \omega_{6}&=\eta_{\mu\nu}\eta_{\rho\sigma}D\wedge P ^{\mu}\wedge  K ^{\nu}\wedge P ^{\rho}\wedge K ^{\sigma}\ .
 \end{align}
 The co-chains $\omega_1$ to $\omega_5$ are closed $(\delta\omega = 0)$, and we therefore have five possible non-trivial co-cycles. However, four of these turn out to be co-boundaries
 \begin{align}
\nonumber
  \omega_{1}&=\frac{1}{4}\delta\Big[\epsilon_{\mu\nu\rho\sigma} P ^{\mu}\wedge  P ^{\nu}\wedge P ^{\rho}\wedge P ^{\sigma}\Big],\\
\nonumber
 \omega_{2}&=\frac{1}{2}\delta\Big [\epsilon_{\mu\nu\rho\sigma}D\wedge P ^{\mu}\wedge  P ^{\nu}\wedge P ^{\rho}\wedge K ^{\sigma}\Big],\\
\nonumber
 \omega_{4}&=-\frac{1}{2}\delta\Big [\epsilon_{\mu\nu\rho\sigma}D\wedge P ^{\mu}\wedge  K ^{\nu}\wedge K ^{\rho}\wedge K ^{\sigma}\Big ],\\
 \omega_{5}&=-\frac{1}{4}\delta\Big [\epsilon_{\mu\nu\rho\sigma}D\wedge K ^{\mu}\wedge  K ^{\nu}\wedge K ^{\rho}\wedge K ^{\sigma}\Big]\ .
 \end{align}
However, it turns out that $\omega_3$ is a non-trivial co-cyle. The only possible $\frak{so}(3,1)$ invariant potential for $\omega_{3}$ would be of the form $\alpha_{3}\sim\epsilon_{\mu\nu\rho\sigma}P^{\mu }\wedge P^{\nu}\wedge K^{\rho}\wedge K^{\sigma}$ but, due to the sign difference between $\delta P^{\mu}$ and $\delta K^{\mu}$, the co-boundary operator annihilates this form, $\delta\alpha_{3}=0$.  Therefore, there is a single non-trivial element of $H^{5}(\mathfrak{so}(4,2),\mathfrak{so}(3,1))$ and correspondingly, a single Wess--Zumino term.
 
The 5-form corresponding to the non-trivial co-cycle $\omega_3$ is given by
 \begin{align}
 \omega_{3}^{\rm wz}&=\epsilon_{\mu\nu\rho\sigma}\,\omega_{D}\wedge\omega_{P}^{\mu}\wedge\omega_{P}^{\nu}\wedge\omega_{K}^{\rho}\wedge\omega_{K}^{\sigma}\\
 &=\epsilon_{\mu\nu\rho\sigma}\Big[e^{4\pi}\left (\xi^{4}\rd \pi\wedge \rd x ^{\mu }\wedge \rd x ^{\nu}\wedge \rd x ^{\rho}\wedge \rd x ^{\sigma}-4\xi^{2}\xi_{\lambda}\rd x ^{\lambda}\wedge \rd x ^{\mu}\wedge \rd x ^{\nu} \wedge \rd x ^{\rho}\wedge \rd x ^{\sigma}\right )\nonumber\\ 
 &\quad +e^{3\pi}\left (-2\xi^{2}\rd \pi\wedge \rd x ^{\mu}\wedge \rd x ^{\nu}\wedge \rd \xi ^{\rho}\wedge \rd x ^{\sigma}+2\xi_{\lambda}\rd x ^{\lambda}\wedge \rd x ^{\mu}\wedge \rd x ^{\nu}\wedge \rd \xi ^{\rho}\wedge \rd \xi ^{\sigma}\right )\nonumber\\
 &\quad +e^{2\pi}\rd \pi\wedge \rd x ^{\mu}\wedge \rd x ^{\nu}\wedge \rd \xi ^{\rho}\wedge \rd \xi ^{\sigma}\Big]\ ,
 \end{align}
 and can be written as a total derivative, 
 \bea && \omega_{3}^{\rm wz}=\rd \beta_{3}^{\rm wz},\nonumber \\
 && \beta_{3}^{\rm wz}=\epsilon_{\mu\nu\rho\sigma}\left[\frac{e^{4\pi}}{4}\xi^{4} \rd x ^{\mu}\wedge  \rd x ^{\nu}\wedge  \rd x ^{\rho}\wedge  \rd x ^{\sigma}-\frac{e^{3\pi}}{3}\xi^{2} \rd x ^{\mu}\wedge  \rd x ^{\nu}\wedge  \rd \xi  ^{\rho}\wedge  \rd x ^{\sigma}+\frac{e^{2\pi}}{2} \rd x ^{\mu}\wedge  \rd x ^{\nu}\wedge  \rd \xi  ^{\rho}\wedge  \rd \xi  ^{\sigma}\right]. \nonumber\\
\eea
 Pulling back and imposing the inverse Higgs constraint \eqref{conformalinversehiggs}, the final result is
 \begin{align}
 S^{\rm wz}_{3}&=\int_{M}\beta_{3}^{\rm wz}= -\frac{1}{2}\int \rd^{4}x\Big[{1\over 2}\square\pi(\partial\pi)^{2}+{1\over 4}(\partial\pi)^{4}\Big]\ , 
 \end{align}
 which reproduces $\mathcal L_3^{\rm}$.

The extension to $d$ space-time dimensions proceeds without too much trouble.  When $d$ is even, there is a single Wess--Zumino galileon, the middle one ${\cal L}_{{d\over 2}+1}$.  The others are all coset constructible.   As an example, in $d=2$ the kinetic term ${\cal L}_2$ is a Wess--Zumino term.  It is impossible to construct with the coset method, since the only possible curvature term which could give it, $R$, is a total derivative in two dimensions.    When $d$ is odd, there is no Wess--Zumino term, and all the conformal galileons are coset constructible. 

It is worth noting that the $4$-dimensional Wess--Zumino term
\be
\mathcal L_3^{\rm } \sim (\partial\pi)^4+2\square\pi(\partial\pi)^2~,
\label{athmwz}
\ee
has been of some interest recently in connection with the a-theorem in four dimensions \cite{Komargodski:2011vj, Komargodski:2011xv}. This term for the $4$ dimensional conformal group plays a similar role to that of the more well-known $2$ dimensional Wess--Zumino term in the trace anomaly.  The extension to $d$ dimensions reflects the fact that there is no anomaly for odd $d$, and in even $d$ it is associated with terms of order $d/2$ in the curvature.

\section{DBI galileons}

The DBI galileons are higher-derivative scalar field theories which non-linearly realize higher dimensional Poincar\'e symmetry and retain second order equations of motion.  In four dimensions, realizing 5-d Poincar\'e, they are 
\begin{align}
\mathcal{L}_{1}&=\pi, \nonumber\\
\mathcal{L}_{2}&=-\sqrt{1+(\partial \pi)^{2}} \ ,\nonumber \\
\mathcal{L}_{3}&=-\left [\Pi\right ]+\gamma^{2}\left [\pi^3\right ] \ ,\nonumber \\
\mathcal{L}_{4}& =-\gamma \left (\left [\Pi \right ]^{2} -\left [\Pi^{2}\right ]\right )-2\gamma^{3}\left (\left [\pi^{4}\right ]-\left [\Pi\right ]\left [\pi^3\right ]\right ) \ ,\nonumber \\
\mathcal{L}_{5}& =-\gamma^{2}\left (\left [\Pi\right ]^{3}+2\left [\Pi^{3}\right ]-3\left [\Pi\right ]\left [\Pi^{2}\right ]\right )-\gamma^{4}\left (6\left [\Pi\right ]\left [\pi^{4}\right ]-6\left [\pi^{5}\right ]-3\left (\left [\Pi\right ]^{2}-\left [\Pi^{2}\right ]\right )\left [\pi^3\right ]\right ) \ .
\end{align}
The notation is explained below \eqref{DBIgalderivexpan}, and 
\be \gamma\equiv {1\over \sqrt{1+(\partial \pi)^2}}.\ee

 The DBI galileons were first derived in \cite{deRham:2010eu} from the perspective of a 3-brane probing a flat $5d$ bulk.\footnote{Their spherical solutions are studied in \cite{Goon:2010xh}, and inflationary non-gaussianity in \cite{Mizuno:2010ag}.} The last four DBI galileons are obtained from Lovelock invariants of the induced brane metric and the boundary terms associated to $5d$ Lovelock invariants,
 \begin{align}
\mathcal{L}_{2}&=-\sqrt{-g} \ ,\nonumber\\
\mathcal{L}_{3}&=\sqrt{-g}K \ ,\nonumber\\
\mathcal{L}_{4}&=-\sqrt{-g}\,R \ , \nonumber\\
\mathcal{L}_{5}&=\sqrt{-g}\,\left[ - K\mn^3+\frac 32 K K\mn^2-\frac 12 K^3-3(R\mn-\frac 12 R g\mn) K^{\mu\nu}
\right] \ ,
\label{branedbigals}
\end{align}
where the induced metric and extrinsic curvature are 
\be g_{\mu\nu}=\eta_{\mu\nu}+\partial_\mu \pi\partial_\nu\pi,\ \ \ ~~~~~~~~~~~K_{\mu\nu}=-\gamma\partial_{\mu}\partial_\nu\pi.\ee
The first term, the tadpole, is not constructed from local terms on the brane, but as the five-dimensional volume bounded by the brane (as discussed in \cite{Goon:2011qf}), and is a Wess--Zumino term, as we will see.

The DBI galileons realize spontaneous breaking of the $5d$ Poincar\'e algebra to its $4d$ Poincar\'e subalgebra,
\be \mathfrak{iso}(4,1)\longrightarrow  \mathfrak{iso}(3,1).\ee
  The broken transformations are translations and rotations into the fifth direction \cite{deRham:2010eu,Goon:2011qf}
\be
\delta_{P_5}\pi=1~,~~~~~~~~~~~~~~~~~~~~~~~\delta_{J_{\mu 5}}\pi=x_\mu+\pi\partial_\mu\pi \ .
\ee
The $5d$ Poincar\'e algebra has the commutation relations
\begin{align}
&\left[J_{MP}, P_Q\right] = \eta_{MQ}P_N-\eta_{NQ}P_M \nonumber\\
&\left[J_{MN}, J_{PQ}\right] = \eta_{MP}J_{NQ}-\eta_{NP}J_{MQ}+\eta_{NQ}J_{MP}-\eta_{MQ}J_{NP}~,
\end{align}
where $\eta_{AB}={\rm diag}\left(-1, 1, 1, 1, 1\right)$. The preserved subalgebra is the Poincar\'e subalgebra generated by $\left(J_{\mu\nu}, P_\rho\right)$, where Greek indices run from $0$ to $3$, acting as in \eqref{confpoincare}.

The broken generators are $P_5$ and $J_{\mu 5}$, and the coset space is 
\be {\rm ISO}(4,2)/{\rm SO}(3,1),\ee
parametrized by\footnote{As in the conformal galileon example \eqref{conformalcosetpar}, this differs slightly from our general expression (\ref{spacetimecoset}), which just amounts to a different choice of parametrization of the coset.}
\be
\tilde V = e^{x\cdot P}e^{\pi P_5}e^{\xi^\alpha J_{\alpha 5}}~.
\ee
From this, we can compute the Maurer--Cartan form \eqref{maurercartanst}
\be
\omega = \tilde V^{-1}\rd \tilde V = \omega_P^\alpha P_\alpha + \omega_{P_5} P_5+ \omega_J^{\alpha} J_{\alpha 5}+\frac{1}{2}\omega_J^{\alpha\beta}J_{\alpha\beta}~,
\ee
where the needed components are
\begin{align}
&\omega_P^\alpha = \rd x^\alpha-\frac{\frac{1}{2}\psi^\alpha\psi_\nu}{1+\frac{\psi^2}{4}}\rd x^\nu+\frac{\psi^\alpha}{1+\frac{\psi^2}{4}}\rd\pi \ ,\\
&\omega_{P_5} = \frac{1-\frac{\psi^2}{4}}{1+\frac{\psi^2}{4}}\rd\pi-\frac{\psi_\mu}{1+\frac{\psi^2}{4}}\rd x^\mu \ ,\\
&\omega_{J}^\alpha = \frac{\rd\psi^\alpha}{1+\frac{\psi^2}{4}}~.
\end{align}
Here, inspired by \cite{Bellucci:2002ji}, we have made the field redefinition  
\be
\psi_\mu \equiv \xi_\mu \frac{\tanh\sqrt\frac{-\xi^2}{4}}{\sqrt\frac{-\xi^2}{4}}~,
\ee
to make the field $\psi$ appear quadratically, which simplifies the expressions. We will not consider the coupling of $\pi$ to matter fields, so the explicit form of $\omega_J^{\mu\nu}$ will not be important.

There is an inverse Higgs constraint, since the commutator of $J_{\mu5}$ with the unbroken translations
\be
\left[P_\mu, J_{\nu5}\right] = -\eta_{\mu\nu}P_5~,
\ee
is proportional to the other unbroken generator $P_5$, so the $\psi_\mu$ field is unphysical and may be eliminated in favor of the $\pi$ by setting
$\omega_{P_5} = 0$, leading to the following relationship between the $\pi$ and $\psi_\mu$ fields
\be
\psi_\mu = \frac{2\partial_\mu\pi}{1+\sqrt{1+(\partial\pi)^2}}~.
\ee
The choice of sign for the square root just leads to an overall sign in front of the Lagrangian, and we will choose the $+$ branch. Using this, we may simplify slightly the expressions for the Maurer--Cartan forms
\begin{align}
\label{dbivielbein}
&\omega_P^\alpha = \left(\delta_\mu^\alpha+\frac{\frac{1}{2}\psi_\mu\psi^\alpha}{1-\frac{\psi^2}{4}}\right)\rd x^\mu  \ ,\\
&\omega_{J}^\alpha = \frac{\rd\psi^\alpha}{1+\frac{\psi^2}{4}}.
\end{align}
The vielbein \eqref{vielbeinextract} and inverse vielbein can be extracted from $\omega_P$,
\be 
e^{~\alpha}_\mu=\delta_\mu^\alpha+\frac{\frac{1}{2}\psi_\mu\psi^\alpha}{1-\frac{\psi^2}{4}},\ \ \ e^\mu_{~\alpha}=\delta^\mu_\alpha-\frac{\frac{1}{2}\psi_\alpha\psi^\mu}{1+\frac{\psi^2}{4}}~.
\ee

In fact, the coset construction is exactly equivalent to the brane construction of \cite{deRham:2010eu}. This can be seen by noting that the induced metric associated to the vielbein (\ref{dbivielbein}) is
\be
g_{\mu\nu} = \eta_{\alpha\beta}e_\mu^{~\alpha}e_\nu^{~\beta} = \eta_{\mu\nu}+\partial_\mu\pi\partial_\nu\pi~.
\ee
and similarly, the covariant derivative \eqref{basiccov} of $\xi$, written with spacetime rather than Lorentz indices, is precisely the extrinsic curvature
\be
\mathcal D_\mu\xi_\nu \equiv e_{\mu}^{~\alpha}\({\omega_{J}}\)_\nu^{\ \beta}\eta_{\alpha\beta}= \gamma\partial_\mu\partial_\nu\pi = -K_{\mu\nu}~.
\ee 
The coset construction then instructs us to make any possible contractions of the objects $\left\{K\mn, g\mn\right\}$ to build invariant actions.

The coset construction is entirely equiavlent to the brane construction of \cite{deRham:2010eu}, because the Gauss--Codazzi relation for a flat bulk
\be R_{\mu\nu\rho\sigma}-K_{\mu\rho}K_{\nu\sigma}+K_{\nu\rho}K_{\mu\sigma}=0~,\ee
 allows us to eliminate the Riemann tensor in favor of the extrinsic curvature. In particular, it is possible to construct all of the terms (\ref{branedbigals}) from the coset construction.\footnote{The DBI terms, save the tadpole $\mathcal L_1$, can also be constructed just by wedging the Maurer--Cartan components together as
\begin{align}
\nonumber
\mathcal L_2 &=  -\frac{1}{4!}\epsilon_{\mu\nu\rho\sigma}~\omega_P^\mu\wedge\omega_P^\nu\wedge\omega_P^\rho\wedge\omega_P^\sigma,\\
\nonumber
\mathcal L_3 &= \frac{1}{3!}\epsilon_{\mu\nu\rho\sigma}~\omega_J^\mu\wedge\omega_P^\nu\wedge\omega_P^\rho\wedge\omega_P^\sigma, \\
\nonumber
\mathcal L_4 &= -\frac{1}{2}\epsilon_{\mu\nu\rho\sigma}\omega_J^\mu\wedge\omega_J^\nu\wedge\omega_P^\rho\wedge\omega_P^\sigma, \\
\nonumber
\mathcal L_5 &= \epsilon_{\mu\nu\rho\sigma}~\omega_J^\mu\wedge\omega_J^\nu\wedge\omega_J^\rho\wedge\omega_P^\sigma~,
\end{align}
and then integrating over the spacetime.
}

Note that---just as in the brane construction---we have failed to construct the tadpole term, $\mathcal L_1 = \pi$, from the coset methods in four dimensions. However, it is possible to construct this tadpole as a Wess--Zumino term by considering the 5-form
\be
\omega_1^{\rm wz} = \epsilon_{\mu\nu\rho\sigma}\omega_{P_5}\wedge\omega_P^\mu\wedge\omega_P^\nu\wedge\omega_P^\rho\wedge\omega_P^\sigma~.
\ee
A fairly straightforward calculation reveals that this 5-form is exact,
\bea &&\omega_1^{\rm wz} = \rd\beta^{\rm wz}_1, \nonumber \\
&& \beta_1^{\rm wz} = \pi\epsilon_{\mu\nu\rho\sigma}\rd x^\mu\wedge\rd x^\nu\wedge\rd x^\rho\wedge\rd x^\sigma~.
\eea
The action given by integrating this 4-form is then
\be
S_1 =\int_M\beta_1^{\rm wz} = \int\rd^4x~\pi~,
\ee
which is the action corresponding to the tadpole Lagrangian ${\cal L}_1$. Therefore we see that the tadpole term is a Wess--Zumino term for spontaneously broken Poincar\'e invariance, in contrast to the other DBI galileon terms.

The DBI galileons are obtainable from the coset construction and so are not Wess--Zumino terms (except for the tadpole term). Taking a small-field limit gives the ordinary galileon terms, indicating that the procedure of contracting the algebra can change which terms are Wess--Zumino. For concreteness, here we derived the DBI galileons in four dimensions, but similar remarks apply in all dimensions: none of the DBI galileons will be Wess--Zumino except for the tadpole.  

The case of higher co-dimensions is more subtle (the DBI galileons for higher co-dimension are discussed in \cite{Hinterbichler:2010xn}), but the extension should not be too difficult. The coset construction used here is not new---there are many examples of authors deriving low-energy effective actions for membranes using non-linear realization techniques, for example \cite{West:2000hr, Chryssomalakos:1999xd, Gomis:2006xw}---but to our knowledge the construction of the full set of DBI galileons from this perspective has not appeared elsewhere in the literature.

Based on the expectation that the brane constructions used in \cite{deRham:2010eu,Hinterbichler:2010xn,Goon:2011qf,Goon:2011uw} are equivalent to the coset construction, we can surmise that the DBI-like galileons living on (A)dS and flat spaces and realizing higher dimensional (A)dS and Poincar\'e symmetries, catalogued in \cite{Goon:2011qf,Goon:2011uw} (before taking any small field limits), have the same Wess--Zumino properties as the original DBI galileons studied in this section, that is, the tadpole is Wess--Zumino and the higher order galileons are not.

\section{Conclusions}

We have demonstrated that galileons arise as Wess--Zumino terms for spontaneously broken space-time symmetries.  Their existence is linked to the existence of non-trivial co-cycles in relative Lie algebra cohomology. The galileon terms are the $d$-form potentials for the $(d+1)$-form non-trivial co-cycles.  The existence of the galileons is due to the local algebraic properties of the relevant groups.

We have also used the techniques of non-linear realizations to address multi-galileon theories, showing that they too are Wess--Zumino terms.  Finally, we considered the DBI galileons, showing that they are not Wess--Zumino terms (except for the tadpole term), and we considered the conformal galileons, showing that only the middle conformal galileon is a Wess--Zumino term. 

The simplest example of a galileon theory is the free non-relativistic point particle. Indeed, this case fits into the scheme presented here, since both the tadpole term and the free particle kinetic term are Wess--Zumino terms in the same sense as the more familiar four-dimensional galileons.

The fact that galileons arise due to local algebraic properties is somewhat tantalizing---it is well-known that there is a non-renormalization theorem for galileons; they are not renormalized to any loop order in perturbation theory \cite{Luty:2003vm, Hinterbichler:2010xn}. It may be possible that this non-renormalization is tied to the algebraic properties of the galileon terms. A possibly instructive example is that of anomalies---whose existence is similarly forecast by algebraic properties \'a la BRST---which also have a non-renormalization theorem, although of a slightly different type (anomalies are not renormalized past 1-loop). This raises the possibility that the non-renormalization of galileons may be understood in terms of some deeper topological or algebraic context based upon their construction as Wess--Zumino terms, but unlike the Wess--Zumino--Witten term of the chiral Lagrangian (which are not renormalized due to a quantization condition on their coefficients), there does not appear to be an obvious global topological condition requiring the coefficients of the galileon terms to be quantized.

 \vspace{1cm}

\noindent
{\bf Acknowledgements:} It is our pleasure to thank Ron Donagi, Randy Kamien, Justin Khoury and particularly Gary Gibbons for helpful discussions. This work was supported in part by Department of Energy grant DE-FG05-95ER40893-A020, NSF grant PHY-0930521, NASA ATP grant NNX08AH27G, NASA ATP grant NNX11AI95G, and by the Fay R. and Eugene L. Langberg chair.

\end{document}